\documentclass[journal]{IEEEtran}
\usepackage{amsmath,amssymb,amsthm,graphicx}
\usepackage{cite}
\usepackage[tight,footnotesize]{subfigure}
\usepackage{ifthen}
\usepackage[usenames,dvipsnames]{color}

\ifCLASSINFOpdf
\else
\fi
\theoremstyle{plain}

\newtheorem{Theorem}{Theorem}

\newtheorem{Lemma}{Lemma}

\theoremstyle{remark}
\newtheorem{Remark}{Remark}

\theoremstyle{definition}
\newtheorem{condition}{Condition}

\def\ba{\begin{array}}
\def\ea{\end{array}}
\newcommand{\beq}{\begin{equation}}
\newcommand{\eeq}{\end{equation}}
\newcommand{\bq}{\begin{eqnarray}}
\newcommand{\eq}{\end{eqnarray}}
\newcommand{\bqn}{\begin{eqnarray*}}
\newcommand{\eqn}{\end{eqnarray*}}
\newcommand{\IEEEbq}{\begin{IEEEeqnarray}{rCl}}
\newcommand{\IEEEeq}{\end{IEEEeqnarray}}
\newcommand{\bee}{\begin{enumerate}}
\newcommand{\eee}{\end{enumerate}}
\newcommand{\bi}{\begin{itemize}}
\newcommand{\ei}{\end{itemize}}

\def \node {\mathcal N}
\def \generator {\mathcal G}
\def \load {\mathcal L}

\begin{document}
\title{Design and Stability of Load-Side Primary Frequency Control in Power Systems
\thanks{\textbf{IEEE Trans. on Automatic Control, 2014 to appear.}}
}

\author{Changhong~Zhao,~\IEEEmembership{Student~Member,~IEEE,}
        Ufuk~Topcu,~\IEEEmembership{Member,~IEEE,}
        Na~Li,~\IEEEmembership{Member,~IEEE,}
        and~Steven~Low,~\IEEEmembership{Fellow,~IEEE}% <-this % stops a space
\thanks{This work was supported by NSF CNS
award 1312390, NSF NetSE grant CNS 0911041, ARPA-E grant DE-AR0000226, Southern California Edison, National Science Council of Taiwan R.O.C. grant NSC 103-3113-P-008-001, Caltech Resnick Institute, and 
California Energy Commission's Small Grant Program through Grant 57360A/11-16. A preliminary version of this work has appeared in the Proceedings of the $3^\text{rd}$ IEEE International Conference on Smart Grid Communications, 2012 \cite{zhao2012swing}.}%
\thanks{C. Zhao and S. Low are with the Department of Electrical Engineering, California Institute of Technology, Pasadena, CA, 91125 USA (e-mail: czhao@caltech.edu; slow@caltech.edu).}% <-this % stops a space
\thanks{U. Topcu is with the Department of Electrical and Systems Engineering, University of Pennsylvania, Philadelphia, PA, 19104 USA (e-mail: utopcu@seas.
upenn.edu).}% <-this % stops a space
\thanks{N. Li is with the Laboratory for Information and Decision Systems, Massachusetts Institute of Technology, Cambridge, MA, 02139 USA (email: na\_li@mit.edu).}}

\maketitle

\begin{abstract}
We present a systematic method to design ubiquitous continuous fast-acting distributed load control for primary frequency regulation in power networks, by formulating an optimal load control  (OLC) problem where the objective is to minimize the aggregate cost of tracking an operating point subject 
to power balance over the network. We prove that the swing dynamics and the branch power flows, coupled with frequency-based load control, serve as a distributed 
primal-dual algorithm to solve OLC.   We establish the global asymptotic stability of a multimachine network
under such type of load-side primary frequency control.
These results imply that the local frequency deviations at each bus convey exactly the right information about the 
global power imbalance for the loads to make individual decisions that turn out to be globally optimal.
Simulations confirm that the proposed algorithm can rebalance power and resynchronize bus frequencies after a
disturbance with significantly improved transient performance.
\end{abstract}


% \fi
\IEEEpeerreviewmaketitle

\section{Introduction}

\subsection{Motivation}
Frequency control maintains the frequency of a power system tightly around its nominal
value when demand or supply fluctuates.   It is traditionally implemented on the generation side
and consists of three mechanisms that work at different timescales in concert 
\cite{WoodWollenberg1996, Bergen2000, MachowskiBialek2008}.
The primary frequency control operates at a timescale up to low tens of 
seconds and uses a governor to adjust, around a setpoint, the mechanical power input to a generator 
based on the local frequency deviation.   It is called the {droop control} and is completely decentralized.
The primary control can rebalance power and stabilize the frequency    
but does not in itself restore the nominal frequency.   The secondary frequency control
(called automatic generation control) operates at a timescale up to a minute or so and adjusts the
setpoints of governors  in a control area  in a centralized fashion to drive 
the frequency back to its nominal value and the inter-area power flows to their scheduled values.  
Economic dispatch operates at  a timescale of several minutes or up and schedules the output levels 
of generators that are online and the inter-area power flows.
See \cite{kiani2012hierarchical} for a recent hierarchical model of these three mechanisms 
and its stability analysis.
This paper focuses on load participation in the primary frequency control.

The needs and technologies for ubiquitous continuous fast-acting distributed load participation 
in frequency control at different timescales have started to mature in the last decade or so.
The idea however dates back to the late 1970s.  Schweppe \emph{et al.} advocate its deployment to 
``assist or even replace turbine-governed systems and spinning reserve'' \cite{Schweppe1980}. They also propose to use spot prices to incentivize the users to adapt their consumption 
to the true cost of generation at the time of consumption.   
% In contrast to direct load control, this approach
% allows the loads to choose their consumption pattern based on their need and the spot price, attaining with
% the generation a homeostatic equilibrium ``to the benefit of both the utilities and their customers''.  
Remarkably it was emphasized back then that such frequency adaptive loads will ``allow the system to accept more readily a stochastically fluctuating energy source, such as wind or solar 
generation'' \cite{Schweppe1980}.
This point is echoed recently in, e.g., 
%Trudnowski, et al (2006) 
%\cite{Trudnowski2006}, Lu and Hammerstrom (2006)
%\cite{LuHammerstrom2006}, Short et al (2007)  \cite{Short2007}, Donnelly, et al (2010) \cite{donnelly2010frequency},
%Brooks et al (2010) \cite{brooks2010demand}, Callaway and Hiskens (2011) \cite{CallawayHiskens2011}, and
%Molin\'{a}-Garcia et al (2011) \cite{molina2011decentralized},
\cite{Trudnowski2006, LuHammerstrom2006,Short2007, donnelly2010frequency, brooks2010demand, CallawayHiskens2011,molina2011decentralized},
that argue for ``grid-friendly'' appliances, such as refrigerators, water or space heaters, ventilation systems, and air 
conditioners, as well as plug-in electric vehicles to help manage energy imbalance. For further references, 
see \cite{CallawayHiskens2011}.
Simulations in all these studies have consistently shown significant improvement in performance and
reduction in the need for spinning reserves.
The benefit of this approach can thus be substantial as 
 the total capacity of grid-friendly appliances in the U.S. is estimated in \cite{LuHammerstrom2006} to be about
18\% of the peak demand, comparable to the required operating reserve, currently at 13\% of the peak demand.
The feasibility of this approach is confirmed by experiments reported in \cite{donnelly2010frequency} that 
measured the correlation between the frequency at a 230kV transmission substation and the frequencies at 
the 120V wall outlets at various places in a city in Montana.   They show that local frequency measurements
are adequate for loads to participate in primary frequency control as well as in the damping of electromechanical 
oscillations due to inter-area modes of large interconnected systems.

Indeed a small scale demonstration project has been conducted by the Pacific Northwest National Lab during early 
2006 to March 2007 where 200 residential appliances participated in primary frequency control by 
automatically reducing their consumption (e.g, the heating element of a clothes dryer was turned off
while the tumble continued) when the frequency of the household dropped below a threshold (59.95Hz) \cite{Hammerstrom2007}.
Field trials are also carried out in other countries around the globe, e.g., the  U.K. Market Transformation 
Program \cite{UKProgram2008}.    Even though loads do not yet provide second-by-second or minute-by-minute
\emph{continuous} regulation service in any major electricity markets, the survey in \cite{Heffner2007loads} 
finds that they already provide 50\% of the 2,400 MW contingency reserve in ERCOT (Electric Reliability Council of Texas) and 
30\% of dispatched reserve energy (in between continuous reserve and economic dispatch) in the U.K. market.
Long Island Power Authority (LIPA) developed LIPA Edge that provides 24.9 MW of demand reduction 
and 75 MW of spinning reserve by 23,400 loads for peak power management \cite{kirby2003spinning}. 

While there are many simulation studies and field trials of frequency adaptive  load control as discussed above, 
there is not much analytic study that relates the behavior of the loads and the equilibrium and dynamic behavior 
of a multimachine power network.
% (\cite{kiani2012hierarchical} being a notable exception even though our focus is very different).
Indeed this has been recognized, e.g., in \cite{Trudnowski2006,  Hammerstrom2007, UKProgram2008},
 as a major unanswered question that must be resolved before 
 ubiquitous continuous fast-acting distributed load participation in frequency regulation  will 
 become widespread.
Even though classical models for power system dynamics \cite{WoodWollenberg1996, Bergen2000, MachowskiBialek2008}
that focus on the generator control can be adapted to include load adaptation, they do not consider the cost, or disutility, to
the load in participating in primary frequency control, an important aspect of such an approach 
\cite{Schweppe1980, Hammerstrom2007, CallawayHiskens2011, molina2011decentralized}.  

In this paper we present a systematic method to design ubiquitous continuous fast-acting distributed load
control and establish the global asymptotic stability of a multimachine network under this type of primary 
frequency control.   Our approach allows the loads to choose their consumption pattern based on their need 
and the global power imbalance on the network, attaining with the generation what \cite{Schweppe1980} calls
a \emph{homeostatic equilibrium} ``to the benefit of both the utilities and their customers.''
To the best of our knowledge, this is the first network model and analysis of 
load-side primary frequency control.

\subsection{Summary}

Specifically we consider a simple network model described by  linearized
swing dynamics at generator buses,  power flow dynamics
on the branches, and a measure of disutility to users when they participate in primary frequency control.   
At steady state, the frequencies at different buses are synchronized to a common nominal value and the mechanic power is 
balanced with the electric power at each bus. Suppose a small change in power injection occurs
on an arbitrary subset of the buses, causing the bus frequencies to deviate from their nominal value.
We assume the change is small and the DC power flow model is reasonably accurate. 
Instead of adjusting the generators as in the traditional approach, 
how should we adjust the controllable loads in the network to rebalance power in a way
that minimizes the aggregate disutility of these loads?   
We formulate this question as an optimal load control (OLC) problem, which informally takes the form
\IEEEbq\nonumber
\min_d ~ c(d) & \quad \text{subject to} \quad & \text{power rebalance}
\IEEEeq
where $d$ is the demand vector and $c$ measures the disutility to loads in participating in control.   
 Even though neither frequency
nor branch power flows appear in OLC, we will show that frequency deviations emerge as a measure of the cost of 
power imbalance and branch  flow deviations as a measure of frequency asynchronism.   More strikingly the swing
dynamics together with local frequency-based load control serve as a distributed primal-dual algorithm to solve the dual of OLC.
This primal-dual algorithm is globally asymptotically stable, steering the network to the unique global optimal
of OLC.

These results have four important implications.  First the local frequency deviation at each bus conveys
exactly the right information about the global power imbalance for the loads themselves to make local decisions that
turn out to be globally optimal. This allows a completely decentralized solution without explicit communication to or 
among the loads.
Second the global asymptotic stability of the primal-dual algorithm of OLC suggests that ubiquitous continuous 
decentralized load participation in primary frequency control is stable, addressing a question raised in several prior studies, 
e.g. \cite{Schweppe1980, Trudnowski2006,  Hammerstrom2007, UKProgram2008}.
Third we present a ``forward engineering'' perspective where we start with the basic goal of load control and 
derive the frequency-based controller and the swing dynamics as a distributed primal-dual algorithm to solve the dual
of OLC.   In this perspective the controller design mainly boils down to specifying an appropriate optimization problem (OLC).
Fourth the opposite perspective of ``reverse engineering'' is useful as well where, given an appropriate
frequency-based controller design, the network dynamics will converge to a unique equilibrium that \emph{inevitably} 
solves OLC with an objective function that depends on the controller design.    In this sense any frequency adaptation 
implies a certain disutility function of the load that the control implicitly minimizes. 
For instance the linear controller in \cite{Trudnowski2006, donnelly2010frequency} implies a quadratic disutility function 
and hence a quadratic objective in OLC.

Our results confirm that frequency adaptive loads can rebalance power and resynchronize frequency,
just as the droop control of the generators currently does.
They fit well with the emerging layered control architecture advocated in \cite{ilic2007hierarchical}.

\subsection{Our prior work and structure of paper}

In our previous papers \cite{zhao2012frequency, zhao2012fast, zhao2013optimal} we consider a power network 
that is tightly coupled electrically  and can be modeled as a single generator connected to a group of loads.  
A disturbance in generation causes the (single) frequency to deviate from its nominal value.  The goal is to 
adapt loads, using local frequency measurements in the presence of additive noise, to
rebalance power at minimum disutility.  The model for generator dynamics in \cite{zhao2013optimal} is more detailed
than the model in this paper.  
Here we study a network of generator and load buses with branch flows between them and
their local frequencies during transient.  We use a simpler model for individual generators and
focus on the effect of the network structure on frequency-based load control.

The paper is organized as follows. Section \ref{sec:problem} describes a dynamic model of power networks.
Section \ref{sec:kr} formulates OLC as a systematic method to design load-side primary frequency control
and explains how the frequency-based load control and the system 
dynamics serve as a distributed primal-dual algorithm to solve OLC. 
Section \ref{conv_analysis} proves that the network equilibrium is globally asymptotically stable. 
Section \ref{sec:casestudy} reports simulations of the IEEE 68-bus test system that uses a much more
detailed and realistic model than our analytic model.  The simulation results not only confirm
the convergence of the primal-dual algorithm, but also demonstrate significantly better transient performance.
Section \ref{sec:conclusion} concludes the paper.

\section{Network model}
\label{sec:problem}

Let $\mathbb{R}$ denote the set of real numbers and $\mathbb{N}$ denote the set of non-zero natural numbers. For a set $\node$, let $|\node|$ denote its cardinality. 
A variable without a subscript usually denotes a vector with appropriate components,
e.g., $\omega= (\omega_j, j \in \node)\in  \mathbb{R}^{|\node|}$. For $a,b \in \mathbb{R}$, $a \leq b$, the expression $[\cdot]^b_a$ denotes $\max\left\{\min\{\cdot,~b\},~a\right\}$. 
%where $\load (j)$ denotes the set of loads at bus $j$, $\node$ denotes the set of buses and $\mathcal E$ denotes the set of edges/lines in the transmission network. 
% For a vector $a = (a_1, \dots, a_k)$, $a_{-i}$ denotes $(a_1, \dots, a_{i-1}, a_{i+1}, a_k)$. 
For a matrix $A$, let $A^T$ denote its transpose. For a signal $\omega(t)$ of time, let $\dot \omega$ denote its time derivative $\frac{d \omega}{d t}$. 
%Unless otherwise specified, all the power system quantities are given in actual values and units, instead of per unit values.  
%and $A^*$ its complex conjugate transposed.

The power transmission network is described by a graph 
$(\node,\mathcal{E})$ where $\node=\{1,\dots,|\node|\}$ is the set of buses
and $\mathcal{E} \subseteq \node \times \node$ is the set of transmission lines connecting the buses.    
We make the following assumptions:
\footnote{These assumptions are similar to the standard DC approximation except that we do not assume 
the nominal phase angle difference is small across each link.
}
\begin{itemize}
\item The lines $(i,j) \in \mathcal E$ are lossless and characterized by their reactances $ x_{ij}$.
\item The voltage magnitudes $|V_j|$ of buses $j \in \node$ are constants.
% \item The phase angle differences $|\theta_i-\theta_j|$ between adjacent buses are
% 	small.
\item Reactive power injections at the buses and reactive power flows on the lines are ignored.
\end{itemize}
We assume that $(\node,\mathcal{E})$ is directed, with an arbitrary orientation,
so that if $(i,j)\in \mathcal E$ then $(j, i)\not\in \mathcal E$. We use $(i,j)$ and $i\rightarrow j$ interchangeably to denote a link in $\mathcal E$, and use ``$i: i\rightarrow j$'' and ``$k: j\rightarrow k$'' respectively to denote the set of buses $i$ that are 
predecessors of bus $j$ and the set of buses $k$ that are successors of bus $j$. We also assume without loss of generality that
 $(\node, \mathcal{E})$ is connected. 

The network has two types of buses: generator buses and load buses. A generator bus not only has loads, but also 
an AC generator that converts mechanic power into electric power through a rotating prime mover. A load bus 
has only loads but no generator. We assume that the system is three-phase balanced.
For a bus $j \in \node$, its phase \emph{a} voltage at time $t$ is $\sqrt{2} |V_j| \cos(\omega^0 t + \theta_j^0+\Delta \theta_j(t))$
where $\omega^0$ is the nominal frequency, $\theta_j^0$ is the nominal phase angle, and $\Delta \theta_j(t)$ is the time-varying phase angle deviation. The frequency at bus $j$ is defined as $\omega_j := \omega^0 + \Delta \dot \theta_j$, and we call $\Delta \omega_j := \Delta \dot \theta_j$ the frequency deviation at bus $j$. We assume that the frequency deviations $\Delta \omega_j$ are small for all the buses $j\in\node$ and the differences $\Delta\theta_i - \Delta\theta_j$ between phase angle deviations are small across all the links $(i,j) \in \mathcal{E}$. We adopt a standard dynamic model, e.g., in \cite[Sec. 11.4]{Bergen2000}. 

\textbf{Generator buses.}
We assume coherency between the internal and terminal (bus) voltage phase angles of the generator; see our technical report \cite[Sec. VII-C]{zhao2013power} for detailed
 justification.   Then the dynamics on a generator bus $j$ is modeled by the swing equation
\IEEEbq\nonumber
M_j \Delta \dot{ \omega}_j + D_j' \Delta \omega_j  = {P_j^m}' -P_{\text{loss},j}^0-P^e_j
\IEEEeq
where $M_j >0$ is the inertia constant of the generator. The term $D_j'\Delta \omega_j$ with $D_j'>0$ represents the 
{(first-order approximation of)} deviation in generator power loss due to friction  \cite{Bergen2000} from its nominal 
value $P_{\text{loss},j}^0  { :=\left(D_j' \omega^0\right)/2}$. 
Here ${P^m_j}'$ is the mechanic power injection to the generator, and $P^e_j$ is the electric power export of the 
generator, which equals the sum of loads at bus $j$ and the net power injection from bus $j$ to the rest of the network.

In general, load power may depend on both the bus voltage magnitude (which is assumed fixed) and frequency.
We distinguish between three types of loads, {\em frequency-sensitive}, {\em frequency-insensitive but controllable},
and \emph{uncontrollable loads}.
{We assume the power consumptions of frequency-sensitive (e.g., motor-type) loads increase linearly with frequency deviation and model the aggregate power consumption of these loads by $\hat d_j^0 + D_j'' \Delta \omega_j$ with $D_j''>0$, where $\hat d_j^0$ is its nominal value.}
We assume frequency-insensitive loads can be actively controlled and our goal is to design and analyze these
control laws.
Let $d_j$ denote the aggregate power of the controllable (but frequency-insensitive) loads at bus $j$. 
{Finally let $P_j^l$ denote the aggregate power consumption of uncontrollable (constant power) loads at bus $j$ that 
are neither of the above two types of loads; we assume $P_j^l$ may change over time but is pre-specified.} 
Then the electric power $P^e_j$ is the sum of frequency-sensitive loads, 
controllable loads, uncontrollable loads, and the net power injection from bus $j$ to other buses:
\IEEEbq\nonumber
P^e_j    :=  \hat d_j^0 + D_j''\Delta \omega_j+ d_j + P_j^l	 + \sum_{k: j\rightarrow k} P_{jk} - \sum_{i: i\rightarrow j }  P_{ij}
\IEEEeq
where $P_{jk}$ is the branch power flow from bus $j$ to bus $k$.

\begin{figure}
\centering
\includegraphics[height=3.0 cm]{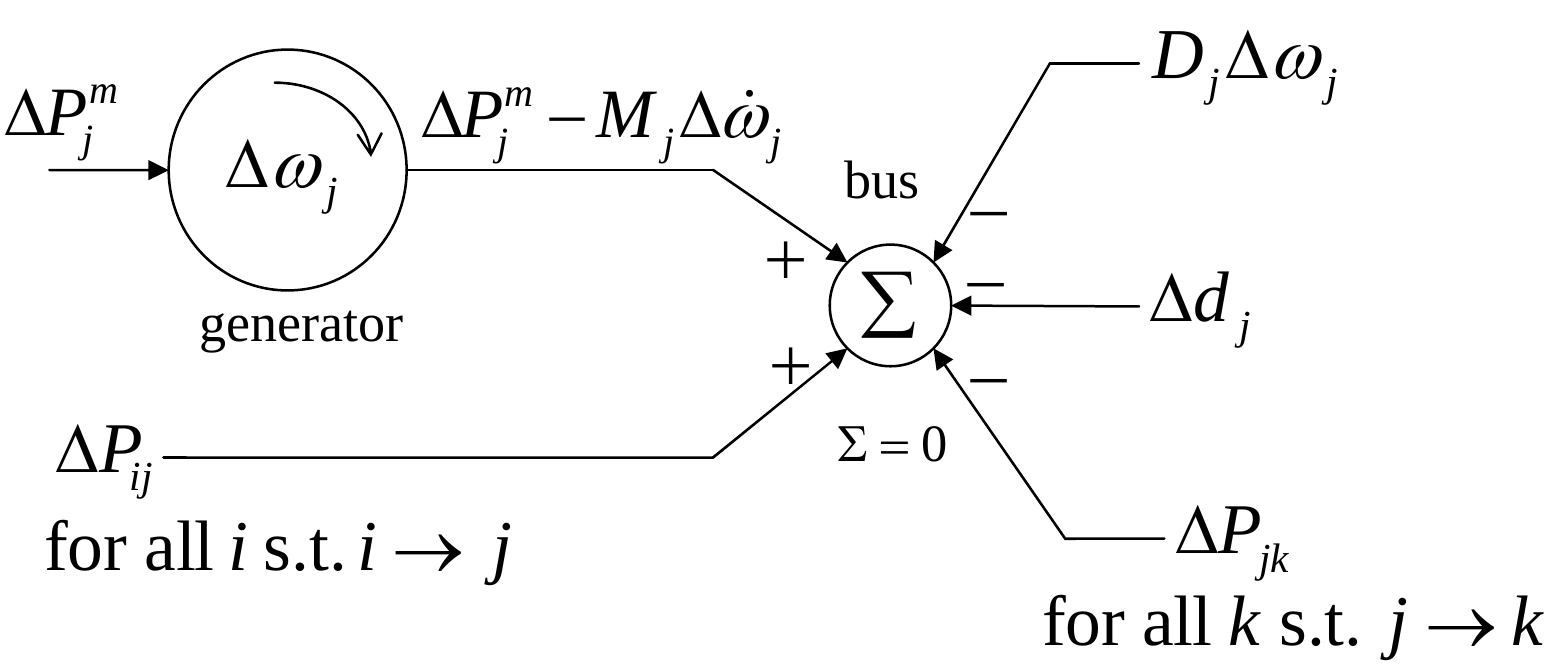}
\caption{Schematic of a generator bus $j$, where $\Delta \omega_j$ is the frequency deviation; $\Delta P^m_j$ is the change in mechanic power minus aggregate uncontrollable load; $D_j \Delta \omega_j$ characterizes the effect of generator friction and frequency-sensitive loads; $\Delta d_j$ is the change in aggregate controllable load; $\Delta P_{ij}$ is the deviation in branch power injected from another bus $i$ to bus $j$; $\Delta P_{jk}$ is the deviation in branch power delivered from bus $j$ to another bus $k$.} \label{fig:bus}
\end{figure}

Hence the dynamics on a generator bus $j$ is
\IEEEbq\nonumber
M_j \Delta \dot{\omega}_j   =   - \left(D_j \Delta \omega_j + d_j  - P_j^m + P_j^{\text{out}} - P_j^{\text{in}} \right)
\IEEEeq
where $D_j := D_j' + D_j''$,  $P_j^m := {P_j^m}' -P_{\text{loss},j}^0 - \hat d_j^0- P_j^l$, 
and $P_j^{\text{out}} := \sum_{k: j\rightarrow k} P_{jk}$ and $P_j^{\text{in}} := \sum_{i: i\rightarrow j} P_{ij}$ are 
respectively the total branch power flows out and into bus $j$. 
{Note that $P_j^l$ is integrated with ${P_j^m}'$ into a single term $P_j^m$, so that any change in power injection, 
whether on the generation side or the load side, is considered a change in $P_j^m$.} 
Let $d_j^0, P_j^{m,0}, P_{ij}^0$ denote the nominal (operating) point at which 
$ d_j^0  -  P_j^{m,0} +  P_j^{\text{out},0} -  P_j^{\text{in},0} =0 $.   Let  
$d_j(t) = d_j^0 + \Delta d_j(t), P_j^m(t) = P_j^{m,0} + \Delta P_j^m(t),
P_{ij}(t) = P_{ij}^0 + \Delta P_{ij}(t)$.  Then the deviations satisfy 
\IEEEbq
 M_j \Delta \dot{\omega}_j  =   - \left(D_j \Delta\omega_j + \Delta d_j  - \Delta P_j^m + \Delta P_j^{\text{out}} - \Delta P_j^{\text{in}} \right). \IEEEeqnarraynumspace
\label{eq:sd1}
\IEEEeq
Fig. \ref{fig:bus} is a schematic of the generator bus model \eqref{eq:sd1}. 

\textbf{Load buses.} 
A load bus that has no generator is modeled by the following algebraic equation that represents power balance
at bus $j$:\footnote{There may be load buses with large
inertia that can be modeled by swing dynamics \eqref{eq:sd1} as proposed in \cite{ilic2010modeling}.  
We will treat them as generator buses mathematically.}
\IEEEbq
0 =  D_j \Delta\omega_j + \Delta d_j  - \Delta P_j^m + \Delta P_j^{\text{out}} - \Delta P_j^{\text{in}} \label{eq:ld}
\IEEEeq
where $\Delta P_j^m$ represents the change in the aggregate uncontrollable load.

\textbf{Branch flows.}
The deviations $\Delta P_{ij}$ from the nominal branch flows follow the (linearized) dynamics
\IEEEbq
\label{eq:fd}
\Delta \dot P_{ij}  =  B_{ij}\left(\Delta \omega_i - \Delta \omega_j \right)
\IEEEeq
where
\IEEEbq
B_{ij}  :=  3\frac{|V_i||V_j|}{x_{ij}}  \cos \left( \theta_i^0 - \theta_j^0 \right)
\label{eq:realflow.2}
\IEEEeq
is a constant determined by the nominal bus voltages and the line reactance. The same model is studied in the literature \cite{WoodWollenberg1996, Bergen2000} based on quasi-steady-state assumptions. 
In \cite[Sec. VII-A]{zhao2013power} we derive this model by solving the differential equation that characterizes 
the dynamics of three-phase instantaneous power flow on reactive lines, without explicitly using quasi-steady-state assumptions. {Note that \eqref{eq:fd} omits the specification of the initial deviations in branch flows $\Delta P(0)$.
In practice $\Delta P(0)$ cannot be an arbitrary vector, but must satisfy
\IEEEbq\label{eq:initial_condition}
\Delta P_{ij}(0) = B_{ij}\left(\Delta \theta_i(0) - \Delta \theta_j (0)\right)
\IEEEeq
for some vector $\Delta \theta (0)$.  In Remark \ref{remark:unique_P} we discuss the implication of this omission on
the convergence analysis.}

%The dynamic model 
%\eqref{eq:fd}--\eqref{eq:realflow.2} is motivated  as follows 
%\cite{Wood_01}, \cite[Chapter 11.4]{Vittal_01}. 
%For balanced three-phase operation at steady state, the voltage angle is $w_0 t + \theta_i^0$
%at bus $i$ (i.e., $v_i(t) = \sqrt{2}|V_i| \cos(w^0t + \theta_i^0))$ and the per-phase branch power is given by 
%\bq
%P_{ij}^0 & = & B_{ij} \sin(\theta_i^0 - \theta_j^0)
%\label{eq:p}
%\eq
%Here  $\theta_i^0$ is the nominal phase angle and $|V_i|$ is the magnitude of 
%the voltage phasor, assumed fixed.
%After a small disturbance the  angle at bus $i$ is perturbed to 
%$\theta_i(t) := \omega^0 t + \theta_i^0 + \Delta\theta_i(t)$.
%We {\em assume} that the branch power is in a quasi-steady state and 
%can be approximated by linearizing \eqref{eq:p} \cite[Chapter 11.4]{Vittal_01} so that,
%to the first order, $P_{ij}(t)  = P_{ij}^0 + B_{ij}  (\Delta\theta_i(t) - \Delta\theta_j(t))$, i.e., 
%\bqn
%\Delta P_{ij}(t) & = & B_{ij}  (\Delta\theta_i(t) - \Delta\theta_j(t))
%\eqn
%Then \eqref{eq:flow_dynamics} follows from $\Delta\dot\theta_j= \Delta\omega_j$.\footnote{While 
%model \eqref{eq:flow_dynamics} assumes that the deviations $\Delta \theta_i$ are small, it does not 
%assume the differences $\theta_i^0 - \theta_j^0$ of their nominal values are small.}

\textbf{Dynamic network model.}  We denote the set of generator buses by $\generator$,  the set of load buses by $\load$, 
and use $|\generator|$ and $|\load|$ to denote the number of generator buses and load buses respectively. 
Without loss of generality label the generator buses so that $\generator=\{1,...,|\generator|\}$ and the
load buses so that $\load=\{|\generator|+1,...,|\node|\}$. In summary the dynamic model of the transmission 
network is specified by (\ref{eq:sd1})--\eqref{eq:fd}.   
To simplify notation we drop the $\Delta$ from the variables denoting deviations and write (\ref{eq:sd1})--\eqref{eq:fd} as:
\begin{IEEEeqnarray}{rCll}
\dot{\omega}_j  & = &  - \frac{1}{M_j} (D_j \omega_j +  d_j  -  P_j^m +  P_j^{\text{out}} -  P_j^{\text{in}} ),~& \forall j \in {\generator}
\label{eq:swing}
\\
0 & = & D_j \omega_j +  d_j  -  P_j^m +  P_j^{\text{out}} -  P_j^{\text{in}},\quad& \forall j \in {\load}
\label{eq:loadbus}
\\
 \dot P_{ij} & = & B_{ij}\left( \omega_i -  \omega_j \right) ,\quad&\forall (i,j) \in {\mathcal{E}}\IEEEeqnarraynumspace
\label{eq:flow_dynamics}
\IEEEeq
where $B_{ij}$ are given by \eqref{eq:realflow.2}.
Hence for the rest of this paper all variables represent {\em deviations} from their nominal values. We will refer to the term $D_j\omega_j$ as the deviation in the (aggregate) frequency-sensitive load even 
though it also includes the deviation in generator power loss due to friction.
{We will refer to $P_j^m$ as a disturbance whether it is in generation or load.}

An \emph{equilibrium point} of the dynamic system \eqref{eq:swing}--\eqref{eq:flow_dynamics} is 
a state $(\omega, P)$ where $\dot\omega_j = 0$ for $j \in \generator$ and $\dot P_{ij}=0$ for $(i,j) \in \mathcal{E}$,
i.e., where all power deviations and frequency deviations are constant over time.

\begin{Remark}
	The model \eqref{eq:swing}--\eqref{eq:flow_dynamics} 
	captures the power system behavior at the timescale of seconds. In this paper we only
	consider a step change in generation or load (constant $P^m$), which implies that the model does not 
	include the action of turbine-governor that changes 
	the mechanic power injection in response to frequency deviation to rebalance power.  Nor does it include any
	secondary frequency control mechanism such as automatic generation control that 
	operates at a slower timescale to restore the nominal frequency.  This model therefore explores the feasibility of 
	fast timescale load control as a 	supplement to the turbine-governor mechanism to resynchronize frequency and 
	rebalance power. 
%Our results in Sections \ref{sec:kr} and \ref{sec:casestudy} suggest it is feasible.
\end{Remark}

We use a much more realistic simulation model developed in \cite{rogers2000power, cheung2009power} to validate
our simple analytic model. The detailed simulations can be found in  \cite[Sec. VII]{zhao2013power}.   We 
summarize the key conclusions from those simulations as follows.
\begin{enumerate}
\item In a power network with long transmission lines, the internal and terminal voltage phase angles of a generator swing coherently, i.e., the rotating speed of the generator is almost the same as the frequency at the generator bus even during  transient.

\item Different buses, particularly those that are in different coherent groups \cite{rogers2000power} and far apart in 
electrical distance \cite{ilic2012toward}, may have different local frequencies for a duration similar to the time 
for them to converge to a new equilibrium, as opposed to resynchronizing almost instantaneously to a common system 
frequency which then converges to the equilibrium.   This particular simulation result justifies
a key feature of our analytic model and is included in Appendix \ref{sec:model_details} of this paper.

\item The simulation model and our analytic model exhibit similar transient behaviors and steady state values for bus 
frequencies and branch power flows.  
\end{enumerate}

\section{Design and stability of primary frequency control}
\label{sec:kr}

Suppose a constant disturbance $P^m = (P^m_j, j\in \node)$ is injected to the set $\node$ of buses.
How should we adjust the controllable loads $d_j$ in \eqref{eq:swing}--\eqref{eq:flow_dynamics} to rebalance 
power in a way that minimizes the aggregate disutility of these loads?   
{In general we can design state feedback controllers of the form $d_j(t) := d_j(\omega(t), P(t))$, prove the
feedback system is globally asymptotically stable, and evaluate the aggregate disutility to the loads at the equilibrium point.
Here we take an alternative approach by directly formulating our goal as an optimal load control (OLC) problem 
and derive the feedback controller as a \emph{distributed} algorithm to solve OLC.  }

We now formulate OLC and present our main results.  These results are proved in Section \ref{conv_analysis}.

\subsection{Optimal load control}
\label{subsec:olc}

%\slow{Make consistent the notation for vectors e.g., $x = (x_i, i\in \node)$ vs $x = (x_1, \dots, x_N)$,
%etc.}

The objective function of OLC consists of two costs. First suppose the (aggregate) controllable load at bus $j$ incurs
a cost (disutility) $\tilde c_j(d_j)$ when it is changed by $d_j$.
{Second the frequency deviation $\omega_j$ causes the (aggregate) frequency-sensitive load at bus $j$ to change 
by $\hat d_j := D_j \omega_j$.   For reasons that will become clear later, we assume that this results in a cost
to the frequency-sensitive load that is proportional to the squared frequency deviation weighted by its relative
damping constant:
\begin{IEEEeqnarray}{rCl}\nonumber
\frac{\kappa D_j}{\sum_{i \in \node} D_i}  \omega_j^2  =:  \frac{\kappa}{D_j \left(\sum_{i \in \node} D_i \right)}\hat d_j^2
\end{IEEEeqnarray}
where $\kappa>0$ is a constant.    Hence the total cost is
\begin{IEEEeqnarray}{rCl}\nonumber
\sum_{j \in \node} \left(  \tilde c_j (d_j) + \frac{\kappa}{D_j \left(\sum_{i \in \node} D_i \right)}\hat d_j^2 \right).
\end{IEEEeqnarray}
To simplify notation, we scale the total cost by $\frac{1}{2 \kappa} \sum_{i \in \node} D_i $ without loss of generality
and define $c_j(d_j) :=  \tilde c_j (d_j)  \frac{1}{2 \kappa} \sum_{i \in \node} D_i $.
Then OLC minimizes the total cost over $d$ and $\hat{d}$ while balancing generation and load across 
the network: 

\noindent
\textbf{OLC:}
\IEEEbq
\min_{\underline{d} \leq d \leq \overline{d}, \hat{d}}  & \quad&
	\sum_{j \in \node} \left(  c_j (d_j) + \frac{1}{2 D_j} \hat{d}_j^2 \right)
\label{eq:olc.1}
\\
\text{subject to}    &  \quad&\sum_{j \in \node} \left(  d_j + \hat{d}_j  \right)  =  \sum_{j \in \node} P^m_j
\label{eq:olc.2}
\IEEEeq
where $-\infty < \underline{d}_j \leq \overline{d}_j < \infty$.  

\begin{Remark}
Note that \eqref{eq:olc.2}
does not require the balance of generation and load at each individual bus, but only
balance across the entire network.   This constraint is less restrictive and offers more opportunity to minimize
costs.  Additional constraints can be imposed if it is desirable that certain buses, e.g., in the same control area,
rebalance their own supply and demand, e.g., for economic or regulatory reasons.
\end{Remark}
%
%\subsection{Assumptions}

%The main variables and assumptions are summarized in Table \ref{table1:notations} and below
%for ease of reference:
%\begin{table}[htbp]
%\caption{Notations.}
%\centering
%\begin{tabular}{|| l | p{8cm} ||}
%\hline \hline
%	$P^m$ & vector $(P_1^m, \dots, P_N^m)$ of deviations in mechanic power (generation) \\
%	$P^e$  & vector $(P_1^e, \dots, P_N^e)$ of deviations in (real) bus power injection \\
%	$P$  & vector $(P_{ij}, (i, j) \in \mathcal E)$ of deviations in branch power flows \\
%\hline
%	$\omega$ &  vector $(\omega_1, \dots, \omega_N)$ of bus frequency deviations, in per unit.  \\
%	$\omega^0$ & common nominal (operating) frequency  \\
%\hline
%	$d$ & vector $(d_j, l\in \load(j), j = 1, \dots, N)$ of frequency-insensitive loads	\\
%          %$d_j$ & vector $(d_j, l\in \load(j))$ of frequency-insensitive loads	\\
%	$\hat{d}$ & vector $(\hat{d}_1, \dots, \hat{d}_N)$ of frequency-sensitive loads  \\
%\hline
%	$c_j$  &  cost functions associated with loads $d_j$, $l\in \load(j), j=1,\dots, N$    \\
%\hline \hline
%\end{tabular}
%\label{table1:notations}
%\end{table}

We assume the following condition throughout the paper:
\begin{condition}\label{cond.1}
OLC is feasible.   The cost functions $c_j$ are strictly 
convex and twice continuously differentiable on $\left[ \underline d_j, \overline d_j \right]$. 
\end{condition}
   
The choice of cost functions is based on physical characteristics
of loads and user comfort levels. 
Examples functions can be found for air conditioners in \cite{ramanathan2008framework} and 
plug-in electric vehicles in \cite{ma2010decentralized}.
See, e.g., \cite{kiani2012hierarchical, fahrioglu2000designing, samadi2010optimal} for other 
cost functions that satisfy Condition \ref{cond.1}.

\subsection{Main results}

The objective function of the dual problem of OLC is
\IEEEbq\nonumber
&&\sum_{j \in \node}  \Phi_j(\nu):=  \\
&& ~\sum_{j \in \node} 
\min_{\underline{d}_j \leq d_j \leq \overline{d}_j, \hat{d}_j} 
\left( \  c_j (d_j) - \nu d_j  +  \frac{1}{2 D_j} \hat{d}_j^2 - \nu \hat{d}_j 
	+  \nu P^m_j  \right) \IEEEeqnarraynumspace \nonumber
\IEEEeq
where the minimization can be solved explicitly as
\IEEEbq
\Phi_j(\nu)  :=     c_j (d_j(\nu)) - \nu d_j(\nu)  -  \frac{1}{2} D_j \nu^2   +  \nu P^m_j
\label{eq:defPhi}
\IEEEeq
with
\IEEEbq
d_j(\nu)  :=  \left[ c_j^{'-1}(\nu) \right]_{\underline{d}_j}^{\overline{d}_j}. 
\label{eq:pvar}
\IEEEeq
This objective function has a scalar variable $\nu$ and is not separable across buses $j \in \node$.
Its direct solution hence requires coordination across buses. We propose the following
{\em distributed} version of the dual problem over the vector $\nu :=\left(\nu_j, j \in \node\right)$, where
 each bus $j$ optimizes over its own variable $\nu_j$ which are constrained to be equal at optimality:

\noindent
\textbf{DOLC:}
\IEEEbq\nonumber
\max_{\nu} & \quad& \Phi(\nu) := \sum_{j \in \node} \Phi_j(\nu_j) 
%\label{eq:dolc.1}
\\
\text{subject to}   &\quad & \nu_i = \nu_j, \quad\forall(i,j) \in \mathcal E. \nonumber
%\label{eq:dolc.2}
\IEEEeq

The following two results are proved in Appendices \ref{subsec:prooflemma1}) and \ref{subsec:prooflemma2}). 
Instead of solving OLC directly, they suggest solving DOLC and recovering the
unique optimal point $(d^*, \hat{d}^*)$ of OLC from the unique dual optimal $\nu^*$.
\begin{Lemma}
\label{lemma.1}
The objective function $\Phi$ of DOLC is strictly concave over $\mathbb{R}^{|\node|}$.
\end{Lemma}

\begin{Lemma}
\label{lemma.2}  
\bee
\item DOLC has a unique optimal point $\nu^*$ with $\nu_i^*=\nu_j^* = \nu^*$ for all $i, j \in \node$. \footnote{For simplicity, we abuse the notation and use $\nu^*$ to denote both the vector $\left(\nu_j^*, j \in \node\right)$ and the common value of its components. Its meaning should be clear from the context. }
\item OLC has a unique optimal point $(d^*, \hat{d}^*)$ where $d^*_j = d_j(\nu^*)$ and $\hat{d}_j^* = D_j \nu^*$ for all $j \in \node$.
%\item There is no duality gap between OLC and DOLC.
\eee
\end{Lemma}
To derive a distributed solution for DOLC consider its Lagrangian
\IEEEbq
L(\nu, \pi) :=  \sum_{j \in \node} \Phi_j(\nu_j) - \sum_{(i,j)\in \mathcal E} \pi_{ij} (\nu_i - \nu_j)
\label{eq:defL}
\IEEEeq
where $\nu \in \mathbb{R}^{|\node|}$ is the (vector) variable for DOLC and $\pi \in \mathbb{R}^{|\mathcal E|}$ is the associated dual variable for
the dual of DOLC. Hence $\pi_{ij}$, for all $(i,j)\in \mathcal E$, measure the cost of 
not synchronizing the variables $\nu_i$ and $\nu_j$ across buses $i$ and $j$. Using \eqref{eq:defPhi}--\eqref{eq:defL},
 a partial primal-dual algorithm for DOLC takes the form 
\begin{IEEEeqnarray}{rCll}
\dot{\nu}_j  &=&  \gamma_j \frac{\partial L}{\partial \nu_j}(\nu, \pi)& \nonumber 
\\
&=&- \gamma_j \left(  d_j(\nu_j) + D_j \nu_j - P_j^m + \pi_j^{\text{out}} - \pi_j^{\text{in}} \right),~\forall j&\in \generator   \IEEEeqnarraynumspace
\label{eq:pd.1}
\\
0 &=&\frac{\partial L}{\partial \nu_j}(\nu, \pi) & \nonumber
\\
&=&-\left(d_j(\nu_j) + D_j \nu_j - P_j^m + \pi_j^{\text{out}} - \pi_j^{\text{in}}\right),~~~\forall j&\in \load    \IEEEeqnarraynumspace \label{eq:pd.2}
\\
\dot{\pi}_{ij} &=&  -\xi_{ij} \frac{\partial L}{\partial \pi_{ij}}(\nu, \pi) =  \xi_{ij} (\nu_i - \nu_j), \quad\quad~~~\forall (i,j) &\in \mathcal{E}   \IEEEeqnarraynumspace \label{eq:pd.3}
\IEEEeq
where  $\gamma_j>0$, $\xi_{ij}>0$ are stepsizes and 
$\pi_j^{\text{out}} := \sum_{k: j\rightarrow k} \pi_{jk}$, 
$\pi_j^{\text{in}} := \sum_{i: i\rightarrow j} \pi_{ij}$. 
%Note that for bus $j$, the algorithm on $\nu_j$ takes either \eqref{eq:pd.1} or \eqref{eq:pd.2}, depending on whether $j$ is a generator bus or a load bus.   
We interpret \eqref{eq:pd.1}--\eqref{eq:pd.3} as an algorithm iterating on the
primal variables $\nu$ and dual variables $\pi$ over time $t\geq 0$.
% as follows.   For a generator bus $j$, given the current iterate $(\nu(t), \pi(t))$, one can use $\gamma_j \partial L\left(\nu(t), \pi(t)\right)/\partial \nu_j$ as the rate of change in $\nu_j$.   For a load bus $j$, the current iterate $\nu_j(t)$ is obtained as the solution of 
% \bqn
% \frac{\partial L} {\partial \nu_j} (\nu_j, \nu_{-j}(t), \pi(t)) & = & 0
% \eqn
% given $\pi(t)$ and $\nu_{-j}(t) := (\nu_i(t), i\neq j)$.
%i.e., $\nu_j(t)$ is the maximizer of $L$ given $\pi(t)$ and $\nu_{-j}(t) := (\nu_i(t), i\neq j)$ since $L$ is strictly concave in $\nu$ according to Lemma \ref{lemma.1}.
% For a link $(i,j) \in \mathcal{E}$, given the current iterate $(\nu(t), \pi(t))$, one can use $-\xi_{ij} \partial L\left(\nu(t), \pi(t)\right)/\partial \pi_{ij}$ as the rate of change in $\pi_{ij}$.  
Set the stepsizes to be:
\IEEEbq\nonumber
\gamma_j = M_j^{-1}, & \qquad&  \xi_{ij} = B_{ij}.
% \nu(0) = \omega(0), & &  \pi(0) = P(0),
\IEEEeq
Then \eqref{eq:pd.1}--\eqref{eq:pd.3} become identical to (\ref{eq:swing})--\eqref{eq:flow_dynamics} if 
we identify $\nu$ with $\omega$ and $\pi$ with $P$, and use $d_j(\omega_j)$ defined by
\eqref{eq:pvar} for $d_j$  in \eqref{eq:swing}--\eqref{eq:loadbus}.
This means that the frequency deviations $\omega$ and the branch flows $P$ are respectively the primal and 
dual variables of DOLC, and the network dynamics, together with frequency-based load control, execute
a primal-dual algorithm for DOLC.

%%, i.e., 
%\bqn
%\nu_j = \omega_j, & & \pi_{ij} = P_{ij}.
%\eqn
%
%dual variabls the way $\nu$ and $\pi$ iterate in the primal-dual algorithm \eqref{eq:pd.1}--\eqref{eq:pd.3} follows exactly the same trajectory as the way $\omega$ and $P$ evolve according to power system dynamics (\ref{eq:swing})--\eqref{eq:flow_dynamics}. Frequency deviations $\omega$ and branch flow deviations $P$ respectively play the roles of primal and dual variables for DOLC and its dual.

\begin{Remark}
Note the consistency of units between the following pairs of quantities: 1) $\gamma_j$ and $M_j^{-1}$, 2) $\xi_{ij}$ and $B_{ij}$, 3) $\nu$ and $\omega$, 4) $\pi$ and $P$. Indeed, since the unit of $D_j$ is $\left[\text{watt}\cdot\text{s}\right]$ from \eqref{eq:swing}, the cost \eqref{eq:olc.1} is in $\left[\text{watt}\cdot \text{s}^{-1}\right]$. From \eqref{eq:defPhi} and \eqref{eq:defL}, $\nu$ and $\pi$ are respectively in $\left[\text{s}^{-1}\right]$ (or equivalently $\left[\text{rad}\cdot \text{s}^{-1}\right]$) 
and $\left[\text{watt}\right]$. From \eqref{eq:pd.1}, $\gamma_j$ is in $\left[ \text{watt}^{-1}\cdot \text{s}^{-2}\right]$ which is the same as the unit of $M_j^{-1}$ from \eqref{eq:swing}. From \eqref{eq:pd.3}, $\xi_{ij}$ is in $\left[\text{watt}\right]$ which is the same as the unit of $B_{ij}$ from \eqref{eq:flow_dynamics}.
\end{Remark}

For convenience, we collect here the system dynamics and load control equations:
\begin{IEEEeqnarray}{rCll}
\dot{\omega}_j & = & -\frac{1}{M_j} \left(  d_j + \hat{d}_j - P^m_j 
	+ P_j^{\text{out}} - P_j^{\text{in}}  \right), \quad\forall j&\in \generator  \IEEEeqnarraynumspace \label{eq:1}
\\
0 & = &  d_j + \hat{d}_j - P^m_j 
	+ P_j^{\text{out}} - P_j^{\text{in}} ,\qquad\quad\quad~~ \forall j & \in \load \IEEEeqnarraynumspace
\label{eq:2}
\\
\dot P_{ij} & = & B_{ij}\left(\omega_i - \omega_j \right),\qquad\qquad\qquad\qquad~~ \forall(i,j)&\in \mathcal E
\IEEEeqnarraynumspace\label{eq:3}
\\
\hat{d}_j & =  &  D_j \omega_j , \qquad\qquad\qquad\qquad\qquad\qquad\quad~~\forall j &\in \node
\IEEEeqnarraynumspace\label{eq:4}
\\
d_j & = & \left[ c_j^{'-1}(\omega_j) \right]_{\underline{d}_j}^{\overline{d}_j},\qquad\qquad\qquad\qquad\quad\quad~ \forall j &\in \node.
\IEEEeqnarraynumspace\label{eq:5}
\IEEEeq
The dynamics \eqref{eq:1}--\eqref{eq:4} are automatically carried out by the system while the active control \eqref{eq:5} needs to be implemented at each controllable load.
Let $(d(t), \hat{d}(t), \omega(t), P(t))$ denote a trajectory of (deviations of) controllable loads, frequency-sensitive loads, 
frequencies and branch flows, generated by the dynamics \eqref{eq:1}--\eqref{eq:5} of the load-controlled system. 
\begin{Theorem}
\label{thm.1}
Starting from any $(d(0), \hat{d}(0), \omega(0), P(0))$, every trajectory 
$(d(t), \hat{d}(t), \omega(t), P(t))$ generated by \eqref{eq:1}--\eqref{eq:5} converges 
to a limit $(d^*, \hat{d}^*, \omega^*, P^*)$ as $t \rightarrow \infty$ such that
\bee
\item $(d^*, \hat{d}^*)$ is the unique vector of optimal load control for OLC;
\item $\omega^*$ is the unique vector of optimal frequency deviations for DOLC;
\item $P^*$ is a vector of optimal branch flows for the dual of DOLC.
\eee
\end{Theorem}
We will prove Theorem \ref{thm.1} and its related results in Section \ref{conv_analysis} below. 
%We will henceforth call a point $(d^*, \hat{d}^*, \omega^*, P^*)$ that satisfies the three conditions in Theorem \ref{thm.1} a {\em system optimal}.

\subsection{Implications}

Our main results have several important implications:
\bee

\item \emph{Ubiquitous continuous load-side primary frequency control.}
	Like the generator droop, frequency-adaptive loads can rebalance power and resynchronize
	frequencies after a disturbance.   Theorem \ref{thm.1} implies that a multimachine network under such
	control is globally asymptotically stable.  The load-side control is often faster because of the larger time 
	constants associated with valves and prime movers on the generator side.
	Furthermore OLC explicitly optimizes the aggregate disutility using the cost functions of heterogeneous loads.

\item {\em Complete decentralization.} 
% It has been well known that the common system frequency is a global signal that measures the power imbalance across the {\em entire} network.   
%% The above sentence is not quite right, because if all frequencies are at nominal value, there is no imbalance.
%% Therefore it is the *deviations* from the nominal frequency that contains information. 
%% What is interesting is that deviations at different buses are different.
%%
The local frequency deviations $\omega_j(t)$ at each bus convey exactly the right information about  
global power imbalance for the loads to make local decisions that turn out to be globally optimal. 
This allows a completely decentralized solution without explicit communication among the buses.
	
% \item  {\em Reverse engineering of swing dynamics.}  The system dynamics \eqref{eq:1}--\eqref{eq:4} coupled with the frequency-based load control \eqref{eq:5} serve as a distributed primal-dual algorithm to solve DOLC and its dual problem.
\item {\em Equilibrium frequency.} 
	The frequency deviations $\omega_j(t)$ at all the buses are synchronized to $\omega^*$ at optimality even though they 
	can be different during transient.   However $\omega^*$ at optimality is in general nonzero, implying
	that the new common frequency may be different from the common frequency before the disturbance. 
% Due to the tight frequency regulation limits, the new steady-state frequency deviation should be made small. 
%This objective can be achieved by, e.g., taking relatively small weights for the cost functions $c_j$ in OLC. 
%Indeed, OLC makes it small, as shown by simulations in Section \ref{sec:casestudy}. 
Mechanisms such as isochronous generators \cite{WoodWollenberg1996} or automatic generation control are needed 
to drive the new system frequency to its nominal value, usually through integral action on the frequency deviations.       

\item {\em Frequency and branch flows.} In the context of optimal load control, the frequency deviations $\omega_j(t)$ 
	emerge as the Lagrange multipliers of OLC that measure the cost of power imbalance, whereas the branch
	flow deviations $P_{ij}(t)$ emerge as the Lagrange multipliers of DOLC that measure the cost of frequency 
	asynchronism.   

\item {\em Uniqueness of solution.} Lemma \ref{lemma.2} implies that the optimal frequency deviation $\omega^*$ is unique and hence the optimal load control $(d^*, \hat{d}^*)$ is unique. As shown below, the vector $P^*$ of optimal branch flows is unique if and only if the network is a tree.  Nonetheless Theorem \ref{thm.1} says that, even for a mesh network, any trajectory of branch flows indeed converges to a limit point.  See Remark \ref{remark:unique_P} for further discussion.
\eee

\section{Convergence analysis}\label{conv_analysis}

%Lemma \ref{lemma.2}, proved in Appendix \ref{subsec:prooflemma2}), is simple since Condition \ref{cond.1} guarantees that OLC is a convex problem.  
This section is devoted to the proof of Theorem \ref{thm.1} and other properties as given by Theorems \ref{thm.2} and \ref{thm.3} below. Before going into the details we first sketch out the key 
steps in establishing Theorem \ref{thm.1}, the convergence of the trajectories generated by \eqref{eq:1}--\eqref{eq:5}.
\begin{enumerate}
\item Theorem \ref{thm.2}: 
	The set of optimal points $(\omega^*, P^*)$ of DOLC and its dual and the set of equilibrium points of \eqref{eq:1}--\eqref{eq:5} are nonempty and the same.
	Denote both of them by $Z^*$. 
	
\item Theorem \ref{thm.3}:
	If $(\node, \mathcal{E})$ is a tree network, $Z^*$ is a singleton with a unique equilibrium 
	point $(\omega^*, P^*)$, otherwise (if $(\node, \mathcal{E})$ is a mesh network), $Z^*$ has an uncountably infinite number (a subspace) 
	of equilibria with the same $\omega^*$ but different $P^*$.  

\item Theorem \ref{thm.1}:
We use a Lyapunov argument to prove that every trajectory $(\omega(t), P(t))$ generated by \eqref{eq:1}--\eqref{eq:5} 
approaches a nonempty, compact subset $Z^+$ of $Z^*$ as $t \rightarrow \infty$. Hence, if $(\node, \mathcal{E})$ is a tree network, then Theorem \ref{thm.3} implies that any trajectory $(\omega(t), P(t))$ converges to the unique optimal point $(\omega^*, P^*)$. If $(\node, \mathcal{E})$ is a mesh network, we show with a more careful argument that $(\omega(t), P(t))$ still converges to a point in $Z^+$, as opposed to oscillating around $Z^+$. Theorem 1 then follows from Lemma \ref{lemma.2}.

\end{enumerate}
We now elaborate on these ideas. 

Given $\omega$ the optimal loads $(d, \hat d)$ are uniquely determined by \eqref{eq:4}--\eqref{eq:5}.
Hence we focus on the variables $(\omega, P)$.
Decompose $\omega^T := \left[\omega_{\generator}^T ~~\omega_{\load}^T \right]$ into 
frequency deviations at generator buses and load buses.
%Recall that we assumed the first $|\generator|$ buses $\{1, \dots, |\generator|\}$ are generator buses and the remaining $|\load|$ buses $\{|\generator|+1, \dots, |\node|\}$ are load buses. 
Let ${C}$ be the $|\node| \times |\mathcal E|$ incidence matrix with
$C_{je} = 1$ if $e = (j, k)\in \mathcal E$ for some bus $k\in \node$, $C_{je} = -1$ if
$e = (i,j) \in \mathcal E$ for some bus $i \in \node$, and $C_{je} = 0$ otherwise. We decompose $C$ into an $|\generator| \times |\mathcal E|$ submatrix $C_{\generator}$
corresponding to generator buses
and an $|\load| \times |\mathcal E|$ submatrix $C_{\load}$ corresponding to
load buses, i.e.,
$C= \bigl[\begin{smallmatrix}  	C_{\generator} \\   C_{\load}	\end{smallmatrix}\bigr]$.
Let
\IEEEbq 
\Phi_{\generator} (\omega_{\generator}) :=  \sum_{j \in \generator} \Phi_j(\omega_j),
& \quad &
L_{\generator} (\omega_{\generator}, P)  :=  \Phi_{\generator} (\omega_{\generator})  
- \omega_{ \generator}^TC_{ \generator}P
 \nonumber  \IEEEeqnarraynumspace \\
\Phi_{\load} (\omega_{\load})  := \sum_{j \in \load} \Phi_j(\omega_j),
& \quad  &
L_{\load} (\omega_{\load}, P) := \Phi_{\load}  (\omega_{\load})
 - \omega_{ \load}^TC_{ \load}P. \nonumber  \IEEEeqnarraynumspace
 \IEEEeq
Identifying $\nu$ with $\omega$ and $\pi$ with $P$, we rewrite the Lagrangian for DOLC defined in \eqref{eq:defL}, in terms of $\omega_{\generator}$
and $\omega_{\load}$, as
\IEEEbq
L(\omega, P)   =  \Phi(\omega) - \omega^T C P
 = 
L_{\generator}(\omega_{\generator},P) +L_{\load}(\omega_{\load},P)
. \IEEEeqnarraynumspace \label{eq:defLv}
\IEEEeq
Then \eqref{eq:1}--\eqref{eq:5} (equivalently, \eqref{eq:pd.1}--\eqref{eq:pd.3}) can be rewritten in the vector form as
\IEEEbq
\dot{\omega_{\generator}} & = &  \Gamma_{\generator} \left[\frac{\partial L_{\generator}}{\partial \omega_{\generator}} \left( \omega_{\generator}, P \right)\right]^T \nonumber 
\\
&=&  \Gamma_{\generator} \left( \left[\frac{\partial \Phi_\generator}{\partial \omega_\generator} \left( \omega_\mathcal {G} \right)\right]^T   - C_\generator P \right) ,
\label{eq:1v}
\\
 0& = &  \frac{\partial L_{\load}}{\partial \omega_{\load}} \left( \omega_{\load}, P \right) \ =\  
 \left[ \frac{\partial \Phi_\load}{\partial \omega_\load} \left( \omega_\load \right) \right]^T 
  - C_{\load} P,
\label{eq:2v}
\\
\dot {P} & = &  -\Xi \left[\frac{\partial L}{\partial P} \left( \omega, P \right)\right]^T
	\ =\   \Xi \, C^T \omega
\label{eq:3v}
\IEEEeq
where $\Gamma_{\generator}:= \text{diag} (\gamma_j,~ j \in \generator)$ and $\Xi:= \text{diag} (\xi_{ij}, ~(i,j)\in \mathcal{E})$.
The differential algebraic equations \eqref{eq:1v}--\eqref{eq:3v} describe the dynamics of the power network.

%An $(\omega^*, P^*)$ is called an {\em equilibrium point} (or an {\em equilibrium}) if it satisfies
%\eqref{eq:1v}--\eqref{eq:3v} with $\dot{\omega} = 0$, $\dot{P}=0$.
A pair $(\omega^*, P^*)$ is called a {\em saddle point} of $L$ if 
\IEEEbq
L(\omega, P^*) \leq  L(\omega^*, P^*) \leq \ L(\omega^*, P),\qquad \forall (\omega, P).
\label{eq:saddle}
\IEEEeq
By \cite[Sec. 5.4.2]{boyd2004convex}, $(\omega^*, P^*)$ is primal-dual optimal for DOLC and its dual if and only if it is a saddle point of $L(\omega, P)$. The following theorem establishes the equivalence between the primal-dual optimal points and the equilibrium points of \eqref{eq:1v}--\eqref{eq:3v}.
\begin{Theorem}\label{thm.2}
A point $(\omega^*, P^*)$ is primal-dual optimal for DOLC and its dual if and only if it is an equilibrium point of \eqref{eq:1v}--\eqref{eq:3v}. Moreover, at least one primal-dual optimal point $(\omega^*, P^*)$ exists and $\omega^*$ is unique among all possible points $(\omega^*, P^*)$ that are primal-dual optimal.
\end{Theorem}
\begin{IEEEproof}
Recall that we identified $\nu$ with $\omega$ and $\pi$ with $P$. In DOLC, the objective function $\Phi$ is (strictly) concave over $\mathbb R^{|\node|}$
(by Lemma \ref{lemma.1}), its constraints
are linear, and a finite optimal $\omega^*$ is attained (by Lemma \ref{lemma.2}).   These facts imply that 
there is no duality gap between DOLC and its dual, and there exists a dual optimal point $P^*$
\cite[Sec. 5.2.3]{boyd2004convex}. 
Moreover, $(\omega^*, P^*)$ is optimal for DOLC and its dual if and only if the following Karush-Kuhn-Tucker (KKT) conditions \cite[Sec. 5.5.3]{boyd2004convex} are satisfied:
\IEEEbq
\text{Stationarity:}&\quad&
\frac{\partial \Phi}{\partial \omega}(\omega^*) \ = \  (CP^*)^T \label{eq:kkt_1}
\\
\text{Primal feasibility:}&\quad&
\omega^*_i \ = \ \omega^*_j, \quad\forall (i,j) \in \mathcal E
\label{eq:kkt_2}.
\IEEEeq
%By \eqref{eq:defLv}, the KKT conditions in \eqref{eq:kkt_1}--\eqref{eq:kkt_2} are also the first-order optimality conditions for $\max_\omega L(\omega, P^*)$ and
%$\min_P L(\omega^*, P)$ since $L$ is (strictly) concave in $\omega$ and convex in $P$.   Hence, $(\omega^*, P^*)$ is primal-dual optimal if and only if it is a saddle point of $L$.
On the other hand $(\omega^*, P^*)= (\omega_{\generator}^*, \omega_{\load}^*, P^*)$ is an equilibrium 
point of \eqref{eq:1v}--\eqref{eq:3v} if and only if \eqref{eq:kkt_1}--\eqref{eq:kkt_2} are satisfied.
%\IEEEbq
%\left[\frac{\partial \Phi_\generator}{\partial \omega_\generator} \left( \omega_\generator^* \right)\right]^T  & =& C_\generator P^*  \nonumber
%\\ 
%\left[\frac{\partial \Phi_\load}{\partial \omega_\load} \left( \omega_\load^* \right)\right]^T  &=&  C_\load P^*  \nonumber 
%\\ 
%\Xi C^T \omega^* & =&0 \nonumber
%\IEEEeq
%which are identical to \eqref{eq:kkt_1}--\eqref{eq:kkt_2}.
Hence  $(\omega^*, P^*)$ is primal-dual optimal if and only if it is an equilibrium point of \eqref{eq:1v}--\eqref{eq:3v}.
%By the existence of primal-dual optimal of DOLC and its dual, such $(\omega^*, P^*)$ exists, and 
The uniqueness of $\omega^*$ is given by Lemma \ref{lemma.2}. 
\end{IEEEproof}

From Lemma \ref{lemma.2}, we denote the unique optimal point of DOLC by $\omega^*1_{\node}=\bigl[\begin{smallmatrix}
	\omega^* 1_\generator \\   \omega^* 1_\load	\end{smallmatrix}\bigr]$, where $1_{\node} \in \mathbb{R}^{|\node|}$ , $1_\generator \in \mathbb{R}^{|\generator|}$ and $1_\load \in \mathbb{R}^{|\load|}$ have all their elements equal to $1$. From \eqref{eq:kkt_1}--\eqref{eq:kkt_2}, define the nonempty set of equilibrium points of \eqref{eq:1v}--\eqref{eq:3v} 
	(or equivalently, primal-dual optimal points of DOLC and its dual) as
\IEEEbq
Z^* & := &  \left\{(\omega, P) \left| \right. \omega=\omega^* 1_{\node}, ~C P = \left[\frac{\partial \Phi}{\partial \omega}\left(\omega^* 1_{\node}\right)\right]^T\right\}. \IEEEeqnarraynumspace
\label{eq:defZstar}
\IEEEeq
%We have proved that $Z^*$ is nonempty and all points in $Z^*$ have unique $\omega= \omega^* 1_{\node}$. 
%As we will show in Theorem \ref{thm.3} below, $Z^*$ is a singleton if and only if the network 
%is a tree.  Before that, we first study the solutions of \eqref{eq:1v}--\eqref{eq:3v}.
%
Let $(\omega^* 1_{\node}, P^*)=(\omega^* 1_\generator, \omega^* 1_\load, P^*) \in Z^*$ be 
{\em any} equilibrium point of \eqref{eq:1v}--\eqref{eq:3v}. We consider a candidate Lyapunov function
\IEEEbq\label{eq:Lyapunov}
U(\omega, P) & = &  \frac{1}{2} \left(\omega_{\generator}-\omega^* 1_{\generator}\right)^T\Gamma_{\generator}^{-1}\left(\omega_{\generator}-\omega^* 1_{\generator}\right) \nonumber
\\&&+ \frac{1}{2} \left(P-P^*\right)^T \Xi^{-1} \left(P-P^* \right).
\IEEEeq
Obviously $U(\omega,P) \geq 0$ for all $(\omega,P)$ with equality if and only if 
$\omega_{\generator} = \omega^* 1_{\generator}$ and $P= P^*$. 
We will show below that $\dot U(\omega,P)  \leq 0$ for all $(\omega,P)$, where $\dot U$ denotes the derivative of $U$ over time along the trajectory $\left(\omega(t), P(t)\right)$.

Even though $U$ depends explicitly only  on $\omega_{\generator}$ and $P$, $\dot{U}$
 depends on $\omega_{\load}$ as well through \eqref{eq:3v}.  
However, it will prove convenient to express $\dot{U}$ as a function of only
$\omega_{\generator}$ and $P$.   To this end, 
write \eqref{eq:2v} as $F(\omega_{\load}, P) = 0$.  Then 
$\frac{\partial F}{\partial \omega_{\load}} (\omega_{\load}, P) = 
\frac{\partial^2 \Phi_{\load}}{\partial \omega_{\load}^2} (\omega_{\load})$
is nonsingular for all $(\omega_{\load}, P)$ from the proof of Lemma \ref{lemma.1} in 
Appendix \ref{subsec:prooflemma1}).
By the inverse function theorem \cite{rudin1976principles}, $\omega_{\load}$ can be written as a continuously
differentiable function of $P$, denoted by $\omega_{\load}(P)$, with
\IEEEbq
\frac{\partial \omega_{\load}} {\partial P} (P) =  
\left( \frac{\partial^2 \Phi_{\load}}{\partial \omega_{\load}^2} \left(\omega_{\load}(P)\right) \right)^{-1} C_{\load}.
\label{eq:dwdP}
\IEEEeq
% From  the proof of Lemma \ref{lemma.1}, $\frac{\partial^2 \Phi_{\load}}{\partial \omega_{\load}^2}$
% is diagonal and negative definite, 
%
Then we rewrite $L(\omega,P)$ as a function of $(\omega_{\generator}, P)$ as
\IEEEbq
L(\omega,P) =  L_{\generator}(\omega_{\generator},P) +L_{\load}\left(\omega_{\load}(P), P\right) 
 =:  \tilde L \left(\omega_{\generator}, P\right). \IEEEeqnarraynumspace
\label{eq:defLtilde}
\IEEEeq
We have the following lemma, proved in Appendix \ref{subsec:prooflemma3}), regarding the properties of $\tilde L$. 
\begin{Lemma}
\label{lemma.3}
$\tilde L$ is strictly concave in $\omega_{\generator}$ and convex in $P$.
\end{Lemma}

Rewrite \eqref{eq:1v}--\eqref{eq:3v} as
\IEEEbq
\dot{\omega}_{\generator} & = &  \Gamma_{\generator} \left[\frac{\partial \tilde L}{\partial \omega_{\generator}} \left( \omega_{\generator}, P \right)\right]^T
\label{eq:1v_tilde}
\\
\dot {P} & = &  -\Xi \left[\frac{\partial \tilde L}{\partial P} \left( \omega_{\generator}, P \right)\right]^T.
\label{eq:2v_tilde}
\IEEEeq
Then the derivative of $U$ along any trajectory $(\omega(t), P(t))$ generated by \eqref{eq:1v}--\eqref{eq:3v} is
\IEEEbq
&&\dot U(\omega, P)  
   = \left(\omega_{\generator} -{\omega}^* 1_{\generator}\right)^T \Gamma_{\generator}^{-1} \, \dot{\omega_{\generator}}+\left(P -{P}^*\right)^T\Xi^{-1}\dot{P} 
 \nonumber 
\\
&  =& \frac{\partial \tilde L} {\partial \omega_{\generator}} (\omega_{\generator}, P) \left(\omega_{\generator} -{\omega}^* 1_{\generator}\right) -\frac{\partial \tilde L} {\partial P} (\omega_{\generator}, P) \left(P -{P}^*\right)   \label{eq:dotU.1} 
\\
& \leq & \tilde L\left(\omega_{\generator}, P \right)-\tilde L\left({\omega}^* 1_{\generator}, P \right) + \tilde L(\omega_{\generator}, P^*)  - \tilde L\left(\omega_{\generator}, P \right) 
 \label{eq:dotU.2} 
\\
& = & L\left(\omega_{\generator}, \omega^* 1_{\load}, P^*\right) - \tilde L\left(\omega^* 1_{\generator}, P\right)  \label{eq:dotU.3} \\
& \leq & L\left(\omega^* 1_{\node}, P\right) -   \tilde L\left(\omega^* 1_{\generator}, P\right)  \label{eq:dotU.4} \\
& = & L_{\generator} \left(\omega^* 1_{\generator}, P\right) + L_{\load} \left(\omega^* 1_{\load}, P\right) \nonumber \\
&& - \left[L_{\generator} \left(\omega^* 1_{\generator}, P\right) + L_{\load} \left(\omega_{\load}(P), P\right)\right] \nonumber
 \\
& \leq & 0
\label{eq:dotU.5}
\IEEEeq
where \eqref{eq:dotU.1} follows from \eqref{eq:1v_tilde}--\eqref{eq:2v_tilde},
the inequality in \eqref{eq:dotU.2} results from Lemma \ref{lemma.3}, the equality in \eqref{eq:dotU.3} holds since  $\omega_\load (P^*) 
= \omega^* 1_{\load}$ by \eqref{eq:kkt_1}, the inequality in \eqref{eq:dotU.4} holds since $ L\left(\omega_{\generator}, \omega^* 1_{\load}, P^*\right) \leq  L\left(\omega^* 1_{\node}, P^*\right) \leq L\left(\omega^* 1_{\node}, P\right)$ from the saddle point condition \eqref{eq:saddle}, and the inequality in \eqref{eq:dotU.5} holds since $\omega_{\load}(P)$ is the maximizer of $L_{\load} \left(\cdot, P\right)$ by the concavity of $L_{\load}$ in $\omega_{\load}$.   

The next lemma, proved in Appendix \ref{subsec:prooflemma4}), characterizes the set in which the value of $U$ does not change over time.
\begin{Lemma}
\label{lemma.4}
$\dot{U}(\omega, P) = 0$ if and only if either \eqref{eq:cond.dotU0_2} or \eqref{eq:cond.dotU0_1} holds:
\IEEEbq
\omega_{\generator}  =  \omega^* 1_{\generator} 
& \quad\text{ and } \quad& 
C_{\load} P  =  
\left[\frac{\partial \Phi_\load}{\partial \omega_{\load}}\left(\omega^* 1_{\load}\right)\right]^T  \label{eq:cond.dotU0_2}
\\
\omega_{\generator}  =  \omega^* 1_{\generator} 
&\quad \text{ and }\quad &
\omega_{\load} (P)  = \omega^* 1_{\load}. \label{eq:cond.dotU0_1}
\IEEEeq
\end{Lemma}
Lemma \ref{lemma.4} motivates the definition of the set  
\IEEEbq
&&E  :=   \left\{ \left(\omega, P\right) \ | \ \dot{U}(\omega, P) = 0 \right\}  \nonumber
\\
&&= 
  \left\{(\omega, P) \left | \right. \omega =\omega^* 1_{\node},~
  C_{\load} P =\left[ \frac{\partial \Phi_\load}{\partial \omega_{\load}}\left(\omega^* 1_{\load}\right)\right]^T\right\} \IEEEeqnarraynumspace \label{eq:defE}
\IEEEeq
 in which $\dot U =0$ along any trajectory $\left(\omega(t),P(t)\right)$. The definition of $Z^*$ in \eqref{eq:defZstar} implies that $Z^* \subseteq E$, as shown in Fig. \ref{fig:EqSets}. 
%(other parts of the figure are explained below).
%%%%%%%%%%%%%%%%%%%%%%
\begin{figure}
\centering
\includegraphics[width=0.4\textwidth]{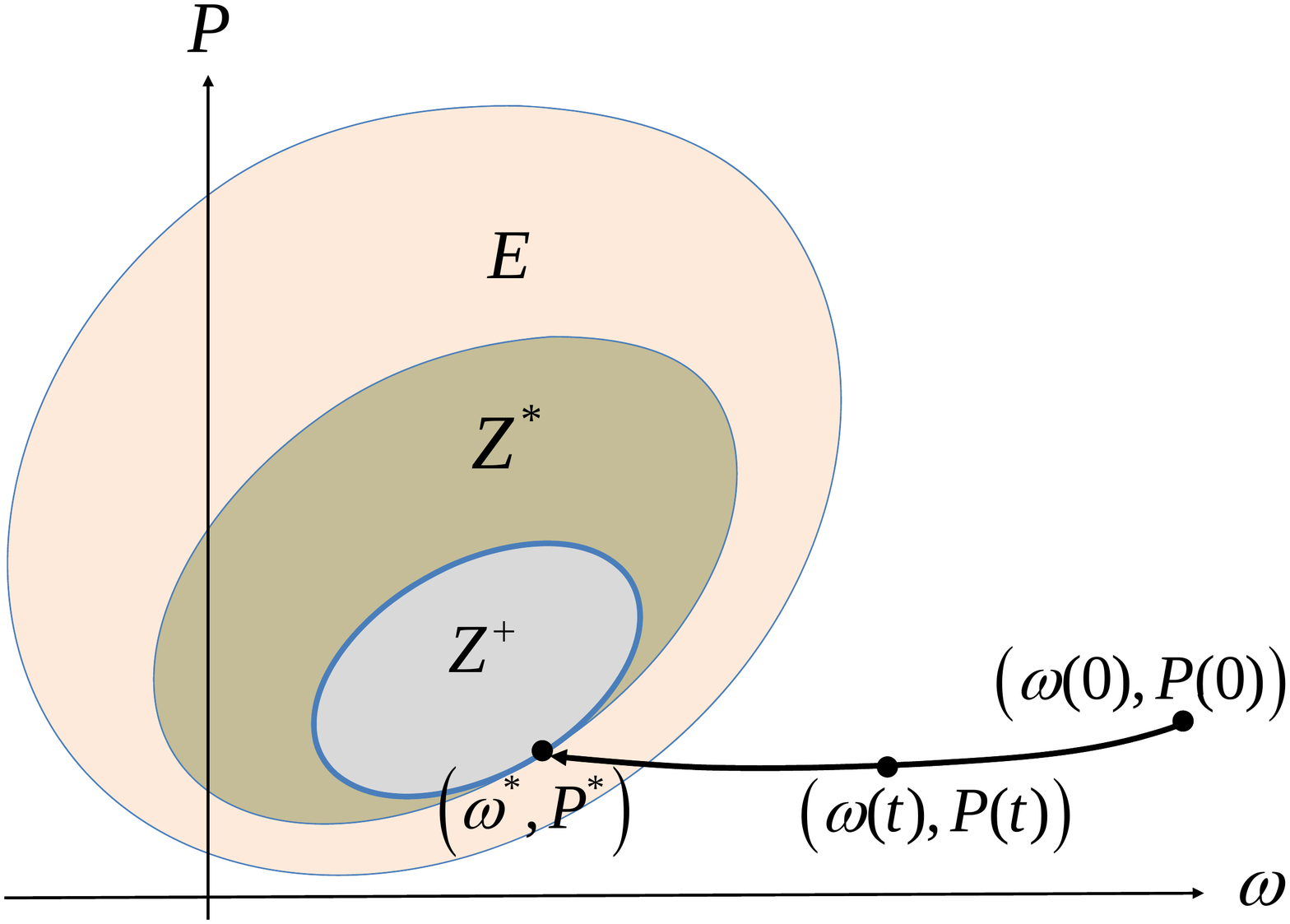}
\caption{$E$ is the set on which $\dot{U} = 0$, 
	$Z^*$ is
	the set of equilibrium points of \eqref{eq:1v}--\eqref{eq:3v}, and $Z^+$ is a compact subset of $Z^*$
	to which all solutions $(\omega(t), P(t))$ approach as $t \rightarrow \infty$. Indeed
	every solution $(\omega(t), P(t))$ converges to a point $(\omega^*,P^*) \in Z^+$ that is
	dependent on the initial state.}
\label{fig:EqSets}
\end{figure}
%%%%%
%%
As shown in the figure $E$ may contain points that are not in $Z^*$. Nonetheless every accumulation point (limit point of any convergent sequence sampled from the trajectory) of a trajectory $(\omega(t), P(t))$ of \eqref{eq:1v}--\eqref{eq:3v} lies in $Z^*$, as the next lemma shows.
\begin{Lemma}
\label{lemma.5}
Every solution $(\omega(t), P(t))$ of \eqref{eq:1v}--\eqref{eq:3v} approaches a nonempty, compact subset (denoted $Z^+$) of $Z^*$ as $t \rightarrow \infty$.
\end{Lemma}
%
%\begin{IEEEproof}
%Since $\dot{U}\leq 0$, the set $\{ (\omega, P) | U( \omega, P) \leq \alpha \}$ is compact and positively
%invariant with respect to \eqref{eq:1v}--\eqref{eq:3v} for any $\alpha$.   Hence, any solution 
%$\left(\omega(t), P(t)\right)$ for $t \geq 0$ stays in the set $\{ (\omega, P) | U( \omega, P) \leq U (\omega(0), P(0)) \}$ and remains bounded.  LaSalle's invariance principle
%then implies that every solution $\left(\omega(t), P(t)\right)$ of \eqref{eq:1v}--\eqref{eq:3v}
%approaches the largest invariant set in $E$ as $t \rightarrow \infty$.  
%Moreover
%\end{IEEEproof}
The proof of Lemma \ref{lemma.5} is given in Appendix \ref{subsec:prooflemma5}). 
The sets $Z^+ \subseteq Z^* \subseteq E$ are illustrated in Fig. \ref{fig:EqSets}. 
Lemma \ref{lemma.5} only guarantees
 that $(\omega(t), P(t))$ approaches $Z^+$ as $t\rightarrow \infty$, while we now show that $(\omega(t), P(t))$ indeed converges to a point in $Z^+$. 
The convergence is immediate in the special case when $Z^*$ is a singleton, but needs a more careful argument when $Z^*$ has multiple points. The next theorem reveals the relation between the number of points in $Z^*$ and the network topology. 
\begin{Theorem}\label{thm.3}

\bee
\item If $(\node,\mathcal E)$ is a tree then $Z^*$ is a singleton.

\item If $(\node,\mathcal E)$ is a mesh (i.e., contains a cycle if regarded as an undirected graph) then $Z^*$ has uncountably many points with the same $\omega^* $ but different $P^*$.  
\eee
\end{Theorem}
\begin{IEEEproof}
From \eqref{eq:defZstar}, the projection of $Z^*$ on the space of $\omega$ is always a singleton $\omega^*1_{\node}$, and hence we only look at the projection of $Z^*$ on the space of $P$, which is  
\IEEEbq\nonumber
Z^*_P :=   \left\{ P \ | \ C P = h^*\right\} 
\IEEEeq
where $h^*:=\left[\frac{\partial \Phi}{\partial \omega}\left({\omega}^*1_{\node}\right)\right]^T$. By Theorem \ref{thm.2}, $Z^*_P$ is nonempty, i.e., there is $P^* \in Z^*_P$ such that $C P^* = h^*$ and hence $1_{\node}^T h^* = 1_{\node}^T C P^* = 0$. Therefore we have
\IEEEbq\label{eq:CPh}
Z^*_P :=  \left\{ P \ | \ \tilde C P = \tilde h^*\right\} 
\IEEEeq
where $\tilde{C}$ is the $(|\node|-1)\times |\mathcal E|$ reduced incidence matrix obtained from $C$ by removing any one of its rows, and $\tilde{h}^*$ is obtained from $h^*$ by removing the corresponding row. Note that $\tilde{C}$ has a full row rank of $|\node|-1$ \cite{mieghem2011graph}.  
If $(\node,\mathcal E)$ is a tree, then $|\mathcal E| = |\node|-1$, so $\tilde{C}$ is square and invertible and $Z_P^*$ is a singleton. If $(\node,\mathcal E)$ is a (connected) mesh, then $|\mathcal E| > |\node|-1$, so $\tilde{C}$ has a nontrivial null space and there are uncountably many points in $Z_P^*$.
\end{IEEEproof}

We can now finish the proof of Theorem \ref{thm.1}.
\begin{IEEEproof}[Proof of Theorem 1] 
For the case in which $(\node,\mathcal E)$ is a tree, Lemma \ref{lemma.5} and Theorem \ref{thm.3}(1) guarantees that every trajectory$(\omega(t), P(t))$ converges to the unique primal-dual optimal point $(\omega^*, P^*)$ of DOLC and its dual, which, by Lemma \ref{lemma.2}, immediately implies Theorem \ref{thm.1}. 

For the case in which $(\node,\mathcal E)$ is a mesh, since $\dot U \leq 0$ along any trajectory $\left(\omega(t), P(t)\right)$, then $U( \omega(t), P(t) ) \leq U (\omega(0), P(0))$ and hence $( \omega(t), P(t) )$ stays in a compact set for $t \geq 0$. Therefore there exists a 
convergent subsequence 
$\left\{(\omega(t_k), P(t_k)),~k \in \mathbb{N}\right\}$, where $0 \leq t_1 < t_2 <...$ and 
$t_k \rightarrow \infty$ as $k \rightarrow \infty$, such that $\lim_{k \rightarrow \infty} \omega(t_k) = \omega^{\infty}$ 
and $\lim_{k \rightarrow \infty} P(t_k) = P^{\infty}$ for some $(\omega^{\infty}, P^{\infty})$.
Lemma \ref{lemma.5} implies that $(\omega^{\infty}, P^{\infty}) \in Z^+ \subseteq Z^*$, and hence $\omega^{\infty} = \omega^* 1_{\node}$ by \eqref{eq:defZstar}. Recall that the Lyapunov function $U$ in \eqref{eq:Lyapunov} can be defined in terms of 
any equilibrium point $(\omega^* 1_{\node}, P^*) \in Z^*$.  In particular,  select $(\omega^* 1_{\node}, P^*) = (\omega^* 1_{\node}, P^{\infty})$, i.e.,
\IEEEbq
U(\omega, P) & := & 
\frac{1}{2} \left(\omega_{\generator}-\omega^* 1_{\generator}\right)^T
	\Gamma_{\generator}^{-1}\left(\omega_{\generator}-\omega^* 1_{\generator}\right)  \nonumber
\\
&& + \frac{1}{2} \left(P-P^\infty \right)^T \Xi^{-1} \left(P-P^\infty \right). \nonumber
\IEEEeq
Since $U \geq 0$ and $\dot U \leq 0$ along any trajectory $\left(\omega(t), P(t)\right)$, 
 $U\left(\omega(t), P(t)\right)$ must converge as ${t \rightarrow \infty}$.   Indeed
 it converges to 0 due to the continuity of $U$ in both $\omega$ and $P$:
 \IEEEbq
 \lim_{t \rightarrow \infty} U\left(\omega(t), P(t)\right)  &=&  
 \lim_{k \rightarrow \infty} U\left(\omega(t_k), P(t_k)\right)  \nonumber
\\
&=& 
U\left(\omega^{\infty}, P^{\infty}\right)  =  0. \nonumber
\IEEEeq
The equation above and the fact that $U$ is quadratic in $(\omega_\generator, P)$ imply that $(\omega_\generator(t), P(t))$ converges to $\left(\omega^* 1_\generator, P^{\infty}\right)$, which further implies that $(\omega(t), P(t))$ converges to $\left(\omega^* 1_\node, P^{\infty}\right)$, a primal-dual optimal point for DOLC and its dual. Theorem \ref{thm.1} then follows from Lemma \ref{lemma.2}.
\end{IEEEproof}

\begin{Remark}
The standard technique of using a Lyapunov function that is quadratic in both the primal and the dual variables 
was first proposed 
by Arrow \emph{et al.} \cite{arrow1958studies}, and has been revisited recently, e.g., in \cite{feijer2010stability, rantzer2009dynamic}.
We apply a variation of this technique to our problem with the following features. 
First, because of the algebraic equation \eqref{eq:2v} in the system, our Lyapunov function is not 
a function of all the primal variables, but only the part $\omega_\mathcal{G}$ corresponding to generator buses. 
Second, in the case of a mesh network when there is a subspace of equilibrium points, we show that the system 
trajectory still converges to one of the equilibrium points instead of oscillating around the equilibrium set.
\end{Remark}

\begin{Remark}\label{remark:unique_P}
Theorems \ref{thm.1}--\ref{thm.3} are based on our analytic model \eqref{eq:1}--\eqref{eq:5} which omits
an important constraint on the initial condition on the branch flows $P(0)$.   As mentioned earlier, in practice,
the initial branch flows must satisfy \eqref{eq:initial_condition} for some $\theta(0)$ (with $\Delta$ dropped). 
With this requirement the branch flow model \eqref{eq:fd}--\eqref{eq:initial_condition} implies $P(t) \in \text{Col}(BC^T)$ for all $t$, where $\text{Col}$ denotes the column space, $B$ is the diagonal matrix with entries $B_{ij}$, and $C$ is the incidence matrix. 
Indeed $P(t) \in \text{Col}(B\tilde C^T)$ since $C^T 1_{\node} =0$ and $\tilde C^T$
with one column from $C^T$ removed has a full column rank. 
A simple derivation from \eqref{eq:CPh} shows that $Z^*_P \cap \text{Col}(B\tilde C^T) = \left\{B\tilde C^T\left(\tilde C B \tilde C^T\right)^{-1} \tilde h^*\right\}$ is a singleton, where $\tilde C B \tilde C^T$ is invertible \cite{mieghem2011graph}. Moreover by \eqref{eq:CPh} and Lemma \ref{lemma.5} we have $P(t) \rightarrow B\tilde C^T\left(\tilde C B \tilde C^T\right)^{-1} \tilde h^*$ as $t \rightarrow \infty$. 
In other words, though for a mesh network the dynamics \eqref{eq:1}--\eqref{eq:5} have a subspace of equilibrium points, all the practical trajectories, whose initial points $\left(\omega(0),P(0)\right)$ satisfy \eqref{eq:initial_condition} for some arbitrary $\theta(0)$, converge to a unique equilibrium point.
\end{Remark}

\section{Case studies}\label{sec:casestudy}

In this section we illustrate the performance of OLC through the simulation of the IEEE 68-bus 
New England/New York interconnection test system \cite{rogers2000power}. The single line diagram of 
the 68-bus system is given in Fig. \ref{fig:ntwk}. 
\begin{figure}
\centering
\includegraphics[height=6.5 cm]{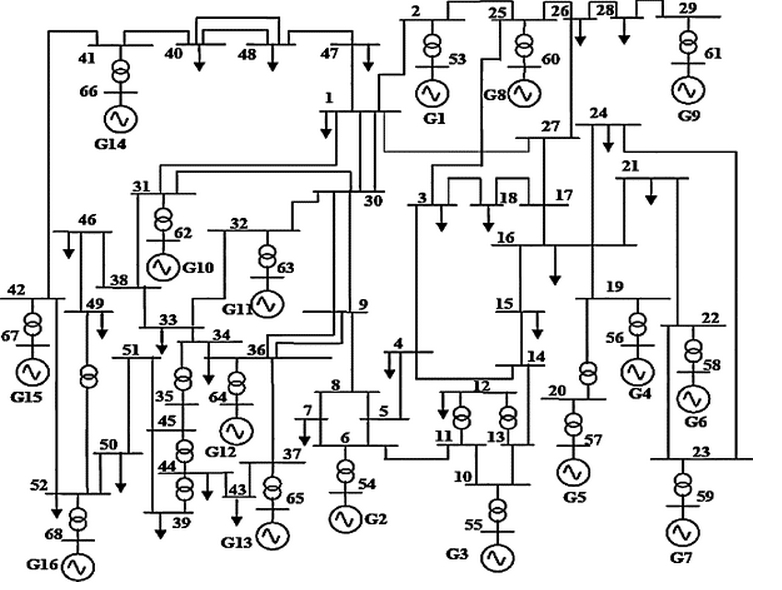}
\caption{Single line diagram of the IEEE 68-bus test system.} \label{fig:ntwk}
\end{figure} 
We run the simulation on Power System Toolbox \cite{cheung2009power}.
Unlike our  analytic model, the simulation model is much more detailed and realistic, 
including two-axis subtransient reactance generator model, IEEE type DC1 exciter model,
 classical power system stabilizer model, AC (nonlinear) power flows, and non-zero line resistances. 
The detail of the simulation model including parameter values can be found in the data files of the toolbox.  
It is shown in \cite{zhao2013power} that our analytic model is a good approximation of the simulation model.

In the test system there are 35 load buses serving different types of loads, including constant active current loads, constant impedance loads, and induction motor loads, with a total real power of 18.23 GW. In addition, we add three loads to buses 1, 7 and 27, each making a step increase of real power by 1 pu (based on 100 MVA), as the $P^m$ in previous analysis. 
We also select 30 load buses to perform OLC. In the simulation we use the same bounds $\left[\underline d, ~\overline d\right]$ with $\underline d = -\overline d$ for each of the 30 controllable loads, and call the value of $30\times \overline d$ the \emph{total size of controllable loads}. We present simulation results below with different sizes of controllable loads. 
The disutility function of controllable load $d_j$ is $c_j(d_j) =  d_j ^2/(2\alpha)$, with identical $\alpha = 100 ~\text{pu}$ 
for all the loads. 
The loads are controlled every $250$ ms, which is a relatively conservative estimate of the rate of load 
control in an existing testbed \cite{douglass2012smart}.              

%Since we have theoretically proved that OLC drives the system to a steady state where the cost of load control is minimized and total generation and total load are balanced, in the simulation 
{We look at the impact of OLC on both the steady state and the transient response of the system, in terms of 
both frequency and voltage.  We present the results with a widely used generation-side stabilizing mechanism known 
as power system stabilizer (PSS) either enabled or disabled. 
Figures \ref{fig:OLC_freq} and \ref{fig:OLC_voltage} respectively show the frequency and voltage at bus 66, under four cases: (i) no PSS, no OLC; (ii) with PSS, no OLC; (iii) no PSS, with OLC; and (iv) with PSS and OLC. In both cases (ii) and (iv), the total size of controllable loads is 1.5 pu. 
%%%%%%%%%%%%%%%%%%%%%%%%%%%%%%%%%%%%%%%
\begin{figure*}[!t]
\centering
\subfigure[]
{\includegraphics[height=5.0 cm]{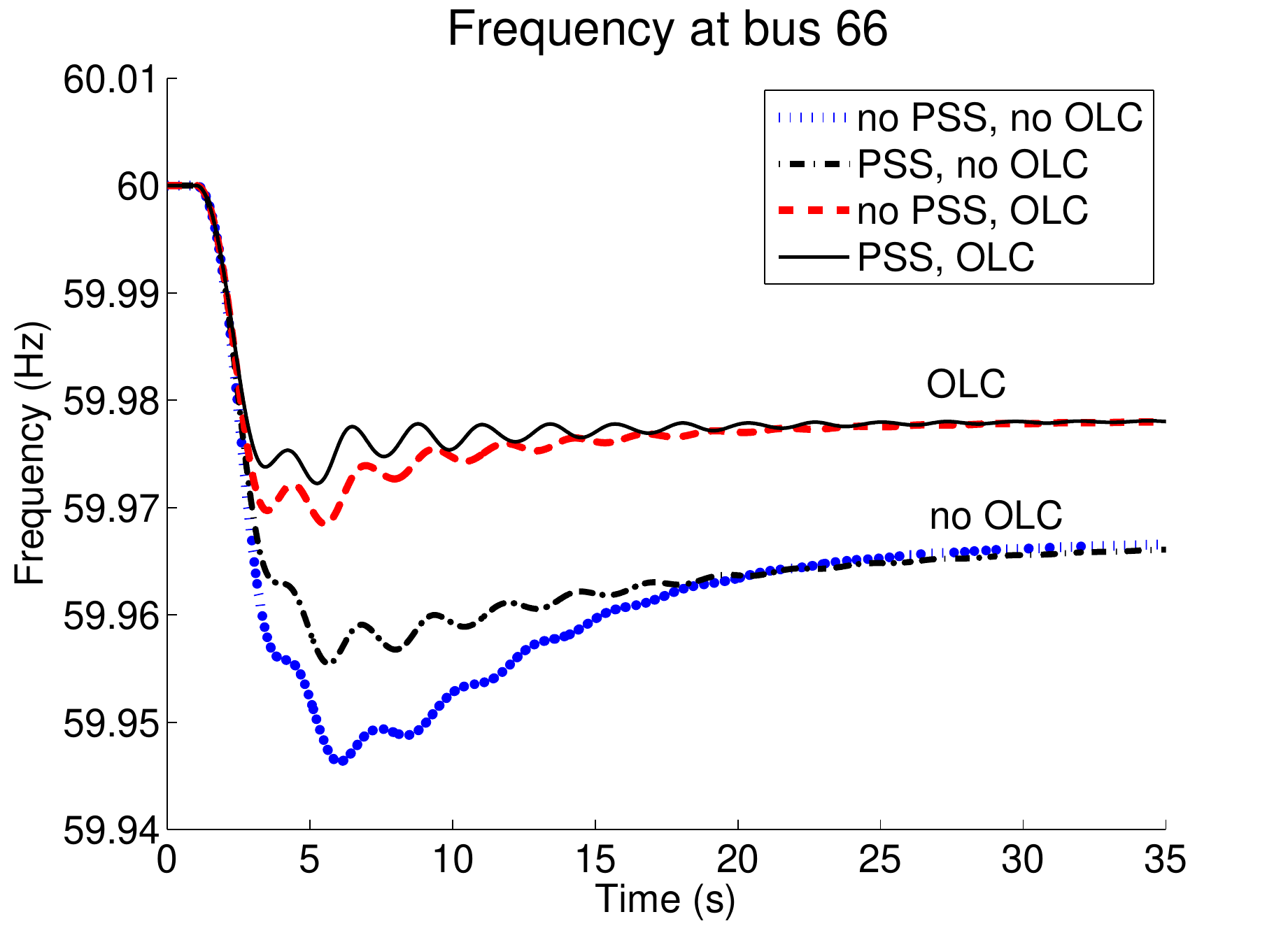}\label{fig:OLC_freq}}
\hfil
\subfigure[]
{\includegraphics[height=5.0 cm]{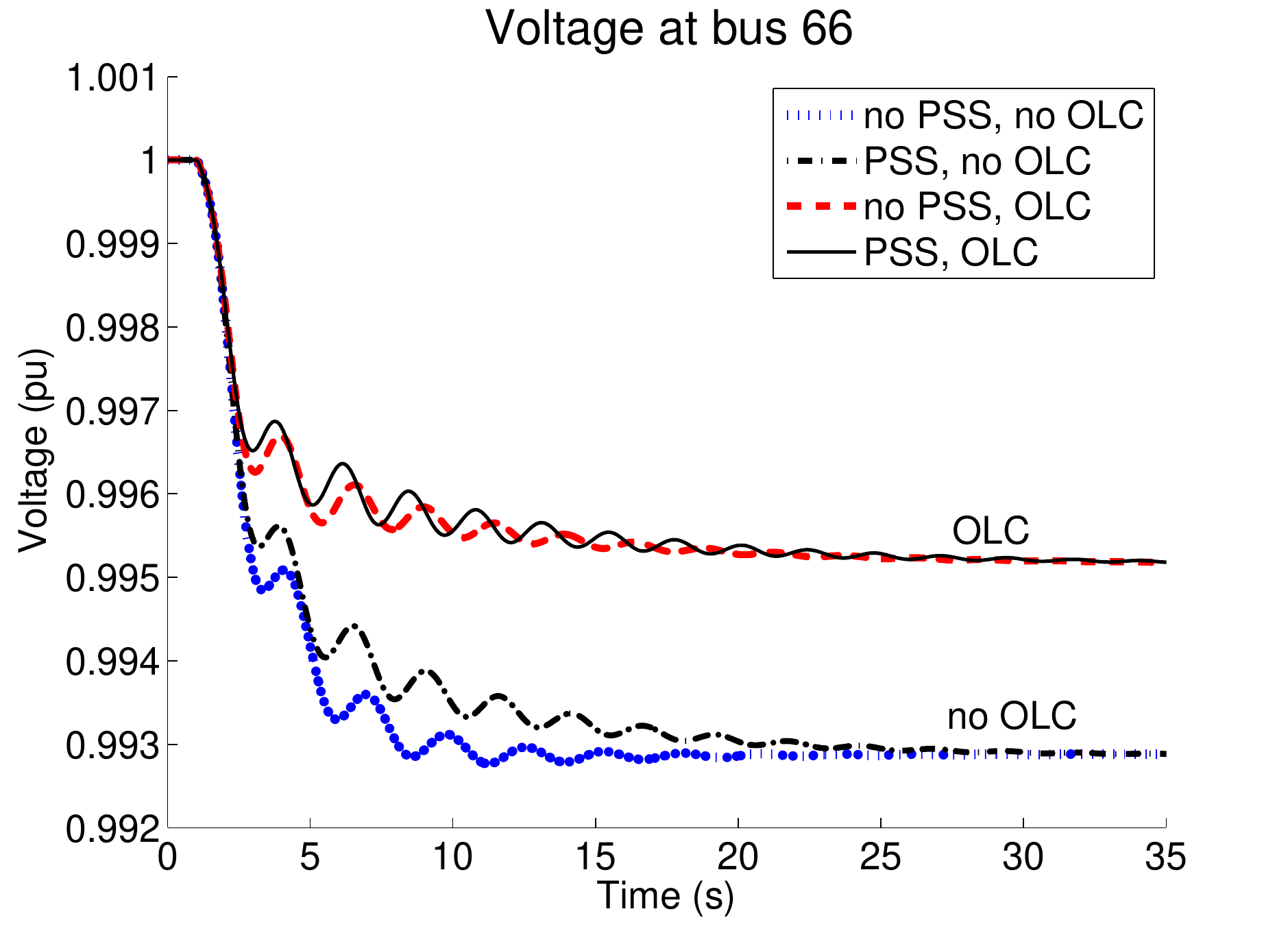}\label{fig:OLC_voltage}}
\caption{The (a) frequency and (b) voltage at bus 66, under four cases: (i) no PSS, no OLC; (ii) with PSS, no OLC; (iii) no PSS, with OLC; (iv) with PSS and OLC.}\label{fig:OLC}
\end{figure*} 
We observe in Fig. \ref{fig:OLC_freq} that whether PSS is used or not, adding OLC always improves the transient response of frequency, in the sense that both the overshoot and the settling time (the time after which the difference between the actual frequency and its new steady-state value never goes beyond $5 \%$ of the difference between its old and new steady-state values) are decreased. Using OLC also results in a smaller steady-state frequency error. Cases (ii) and (iii) suggest that using OLC solely without PSS produces a much better performance than using PSS solely without OLC. 
The impact of OLC on voltage, with and without PSS, is qualitatively demonstrated in Fig. \ref{fig:OLC_voltage}. Similar to its impact on frequency, OLC improves significantly
both the transient and steady-state 
of voltage with or without PSS.  For instance the steady-state voltage is within 4.5\% of the nominal value with OLC
and 7\% without OLC.

To better quantify the performance improvement due to OLC we plot in Figures \ref{fig:ss_size}--\ref{fig:st_size} the new steady-state frequency, the lowest frequency (which indicates overshoot) and the settling time of frequency at bus 66, against the total size of controllable loads. PSS is always enabled.  
\begin{figure*}[!t]
\centering
\subfigure[]
{\includegraphics[height=4.4 cm]{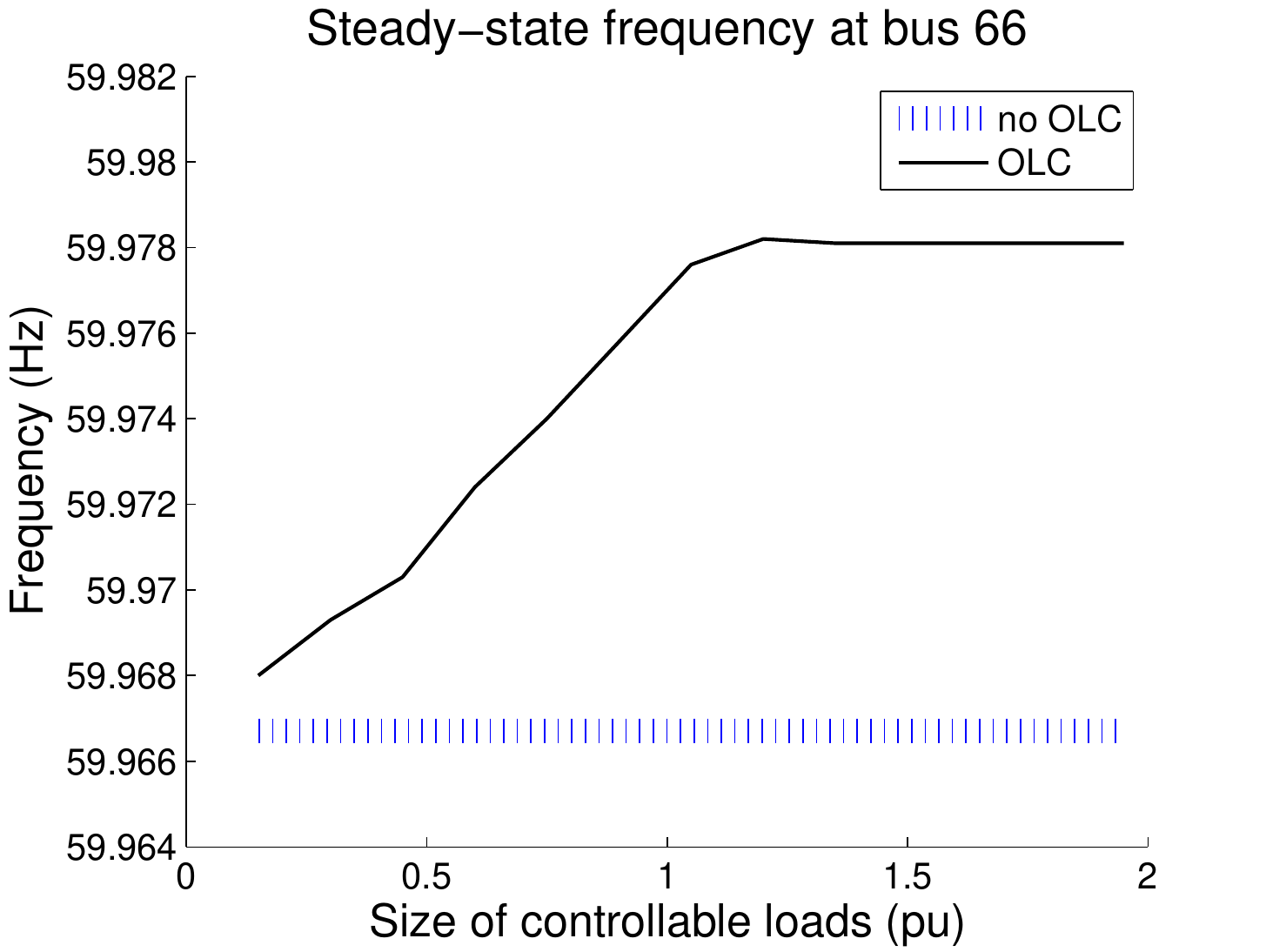}\label{fig:ss_size}}
\hfil
\subfigure[]
{\includegraphics[height=4.4 cm]{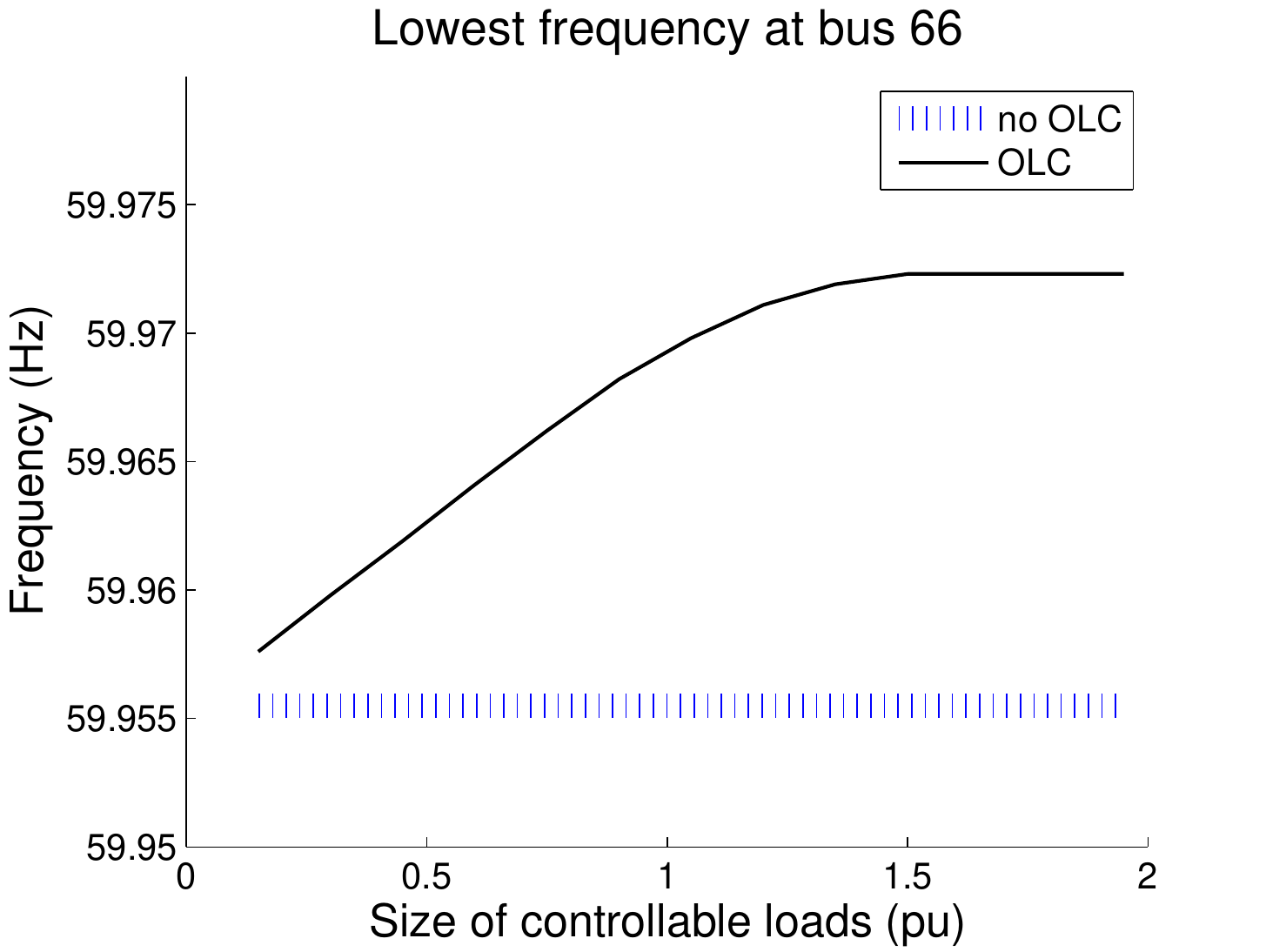}\label{fig:low_size}}
\hfil
\subfigure[]
{\includegraphics[height=4.4 cm]{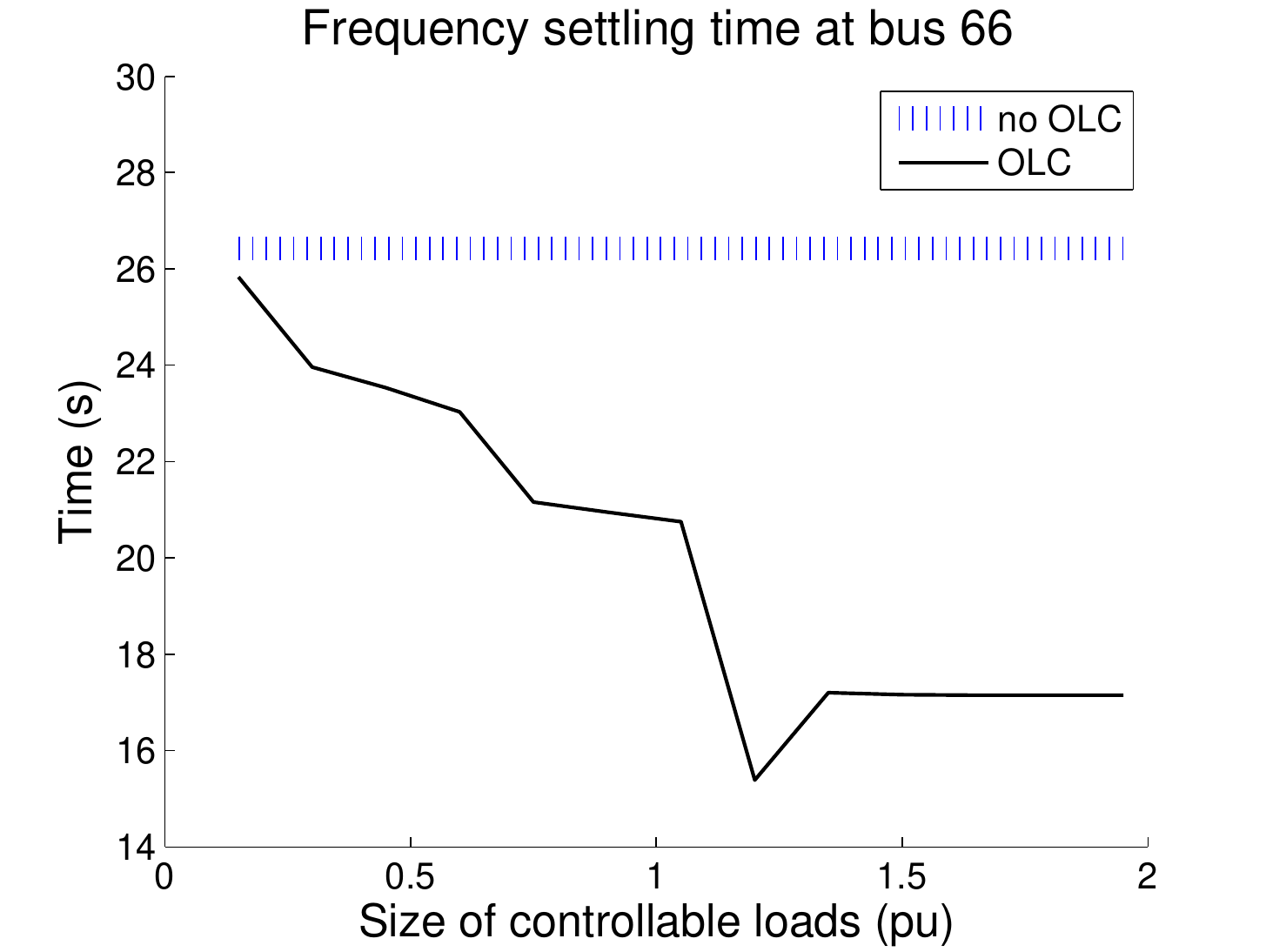}\label{fig:st_size}}
\caption{The (a) new steady-state frequency, (b) lowest frequency and (c) settling time of frequency at bus 66, against the total size of controllable loads.}\label{fig:performance_size}
\end{figure*} 
We observe that using OLC always leads to a higher new steady-state frequency (a smaller steady-state error), a higher lowest frequency (a smaller overshoot), and a shorter settling time, regardless of the total size of controllable loads. As the total size of controllable loads increases, the steady-state error and overshoot decrease almost linearly until a saturation around 1.5 pu. There is a similar trend for the settling time, though the linear dependence is approximate. In summary OLC improves both the steady-state and transient performance of frequency, and in general deploying more controllable loads leads to bigger
improvement.     

{To verify the theoretical result that OLC minimizes the aggregate cost of load control,
Fig. \ref{fig:OLC_disu} shows the cost of OLC over time, obtained by evaluating the quantity
defined in \eqref{eq:olc.1} using the trajectory of controllable and frequency-sensitive loads from the simulation.
We see that the cost indeed converges to the minimum cost for the given change in $P^m$.}   
\begin{figure}
\centering
\includegraphics[height=5.0 cm]{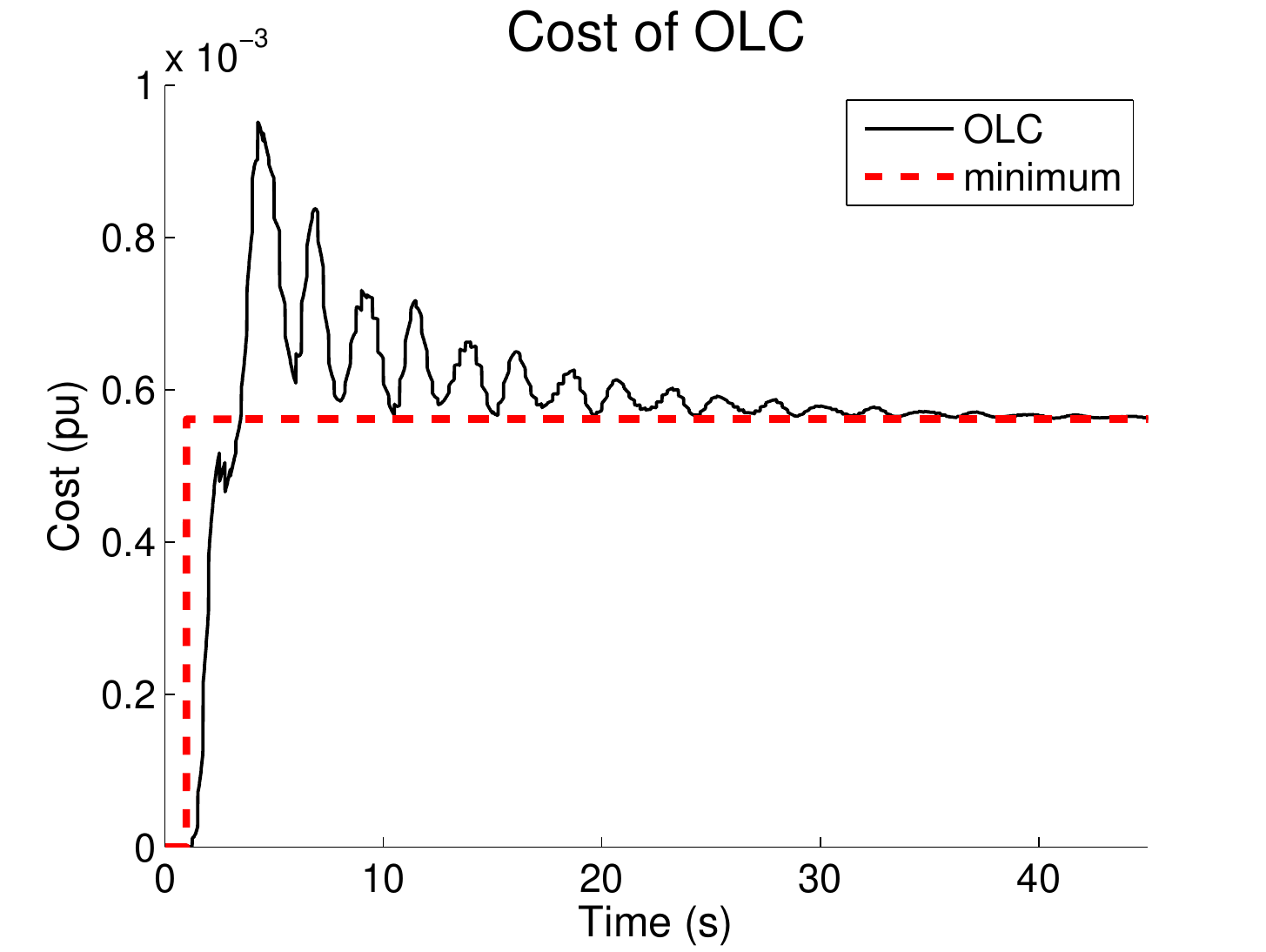}
\caption{The cost trajectory of OLC (solid line) compared to the minimum cost (dashed line).}\label{fig:OLC_disu}
\end{figure}

\section{Conclusion}\label{sec:conclusion}

We have presented a systematic method to design ubiquitous continuous fast-acting distributed load
control for primary frequency regulation in power networks, by formulating an optimal load control 
(OLC) problem where the objective is to minimize the aggregate control cost
subject to power balance across the network. 
We have shown that the dynamics of generator swings and the branch power flows,
 coupled with a frequency-based load control, serve as a distributed primal-dual
algorithm to solve the dual problem of OLC.  
Even though the system has multiple equilibrium points with nonunique branch power flows, we have proved that 
it nonetheless converges to a unique optimal point.  
Simulation of the IEEE 68-bus test system confirmed that the proposed mechanism can rebalance
power and resynchronize bus frequencies 
with significantly improved transient performance.

% or


\appendices
\section{Simulation showing feature of model}\label{sec:model_details}
A key assumption underlying the analytic model \eqref{eq:swing}--\eqref{eq:flow_dynamics} is that different 
buses may have their own frequencies during  transient, instead of resynchronizing almost instantaneously 
to a common system frequency which then converges to an equilibrium. 
Simulation of the 68-bus test system confirms this phenomenon. 
Fig. \ref{fig:group_freq} shows all the 68 bus frequencies from the simulation with the same step change $P^m$ as that in Section \ref{sec:casestudy} but without OLC. To give a clearer view of the 68 bus frequencies, they are divided into the following 4 groups, respectively shown in subfigures \ref{fig:group_freq_1}--\ref{fig:group_freq_4}. 
\begin{enumerate}
\item Group 1 has buses 41, 42, 66, 67, 52, and 68;
\item Group 2 has buses 2, 3, 4, 5, 6, 7, 8, 10, 11, 12, 13, 14, 15, 16, 17, 18, 19, 20, 21, 22, 23, 24, 25, 26, 27, 28, 29, 53, 54, 55, 56, 57, 58, 59, 60, and 61;
\item Group 3 has buses 1, 9, 30, 31, 32, 33, 34, 35, 36, 37, 38, 39, 40, 43, 44, 45, 46, 47, 48, 49, 51, 62, 63, 64, and 65;
\item Group 4 has bus 50 only.
\end{enumerate} 
%%%%%%%%%%%%%%%%%%%%%%%%%%%%%%%%%%%%%%%
\begin{figure*}[!t]
\centering
\subfigure[]
{\includegraphics[height=5.0 cm]{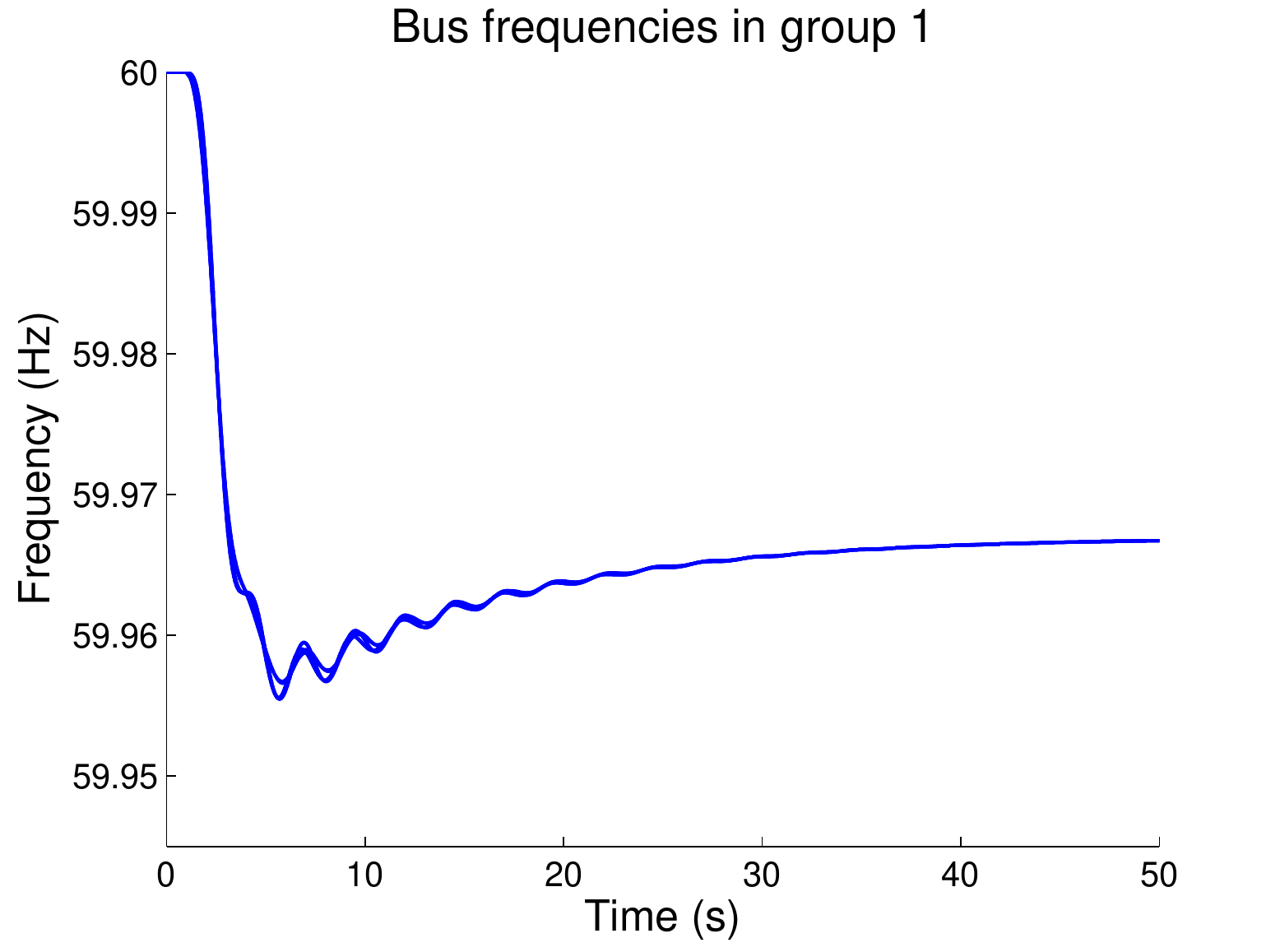}\label{fig:group_freq_1}}
\hfil
\subfigure[]
{\includegraphics[height=5.0 cm]{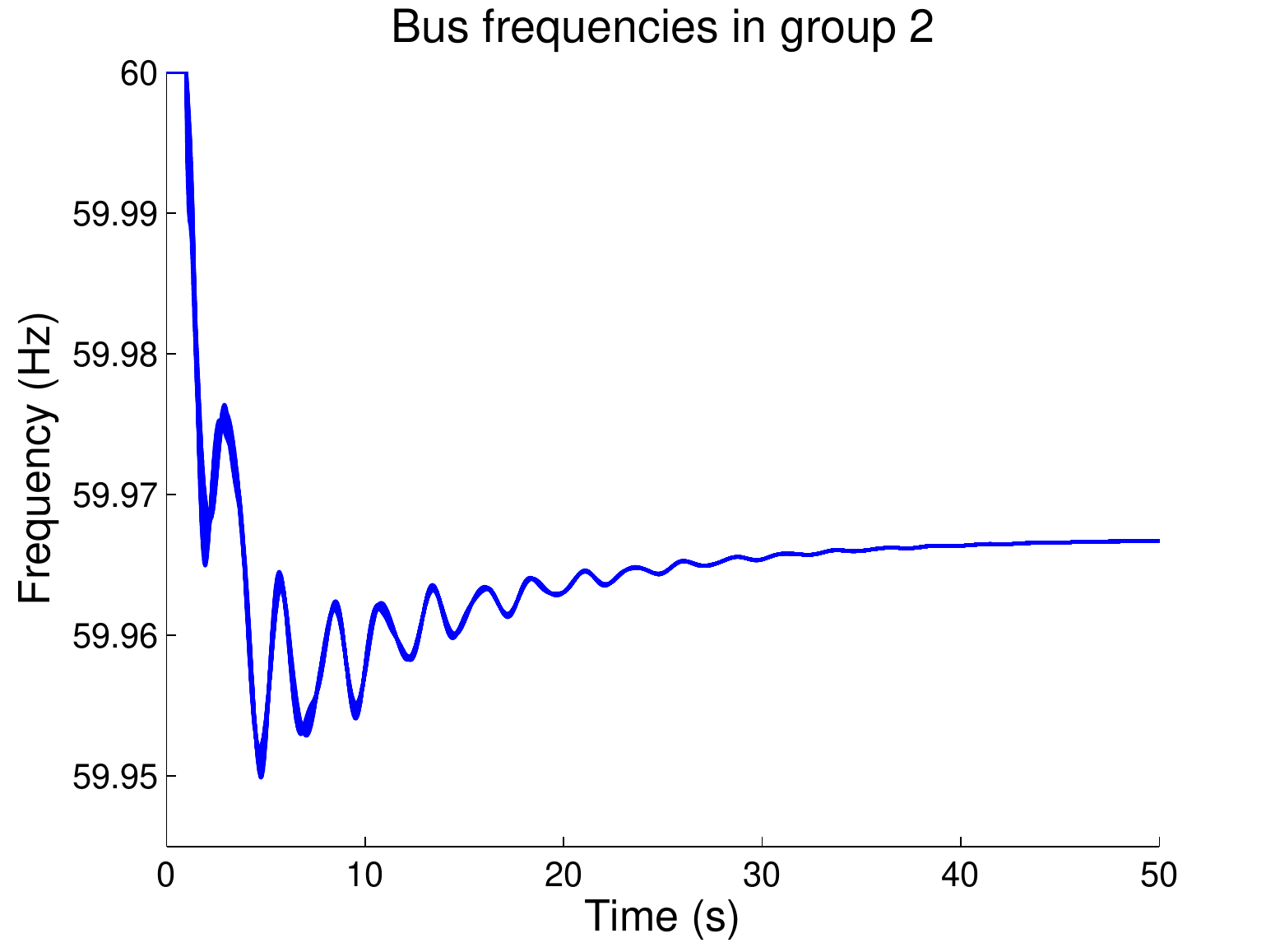}\label{fig:group_freq_2}}
\hfil
\subfigure[]
{\includegraphics[height=5.0 cm]{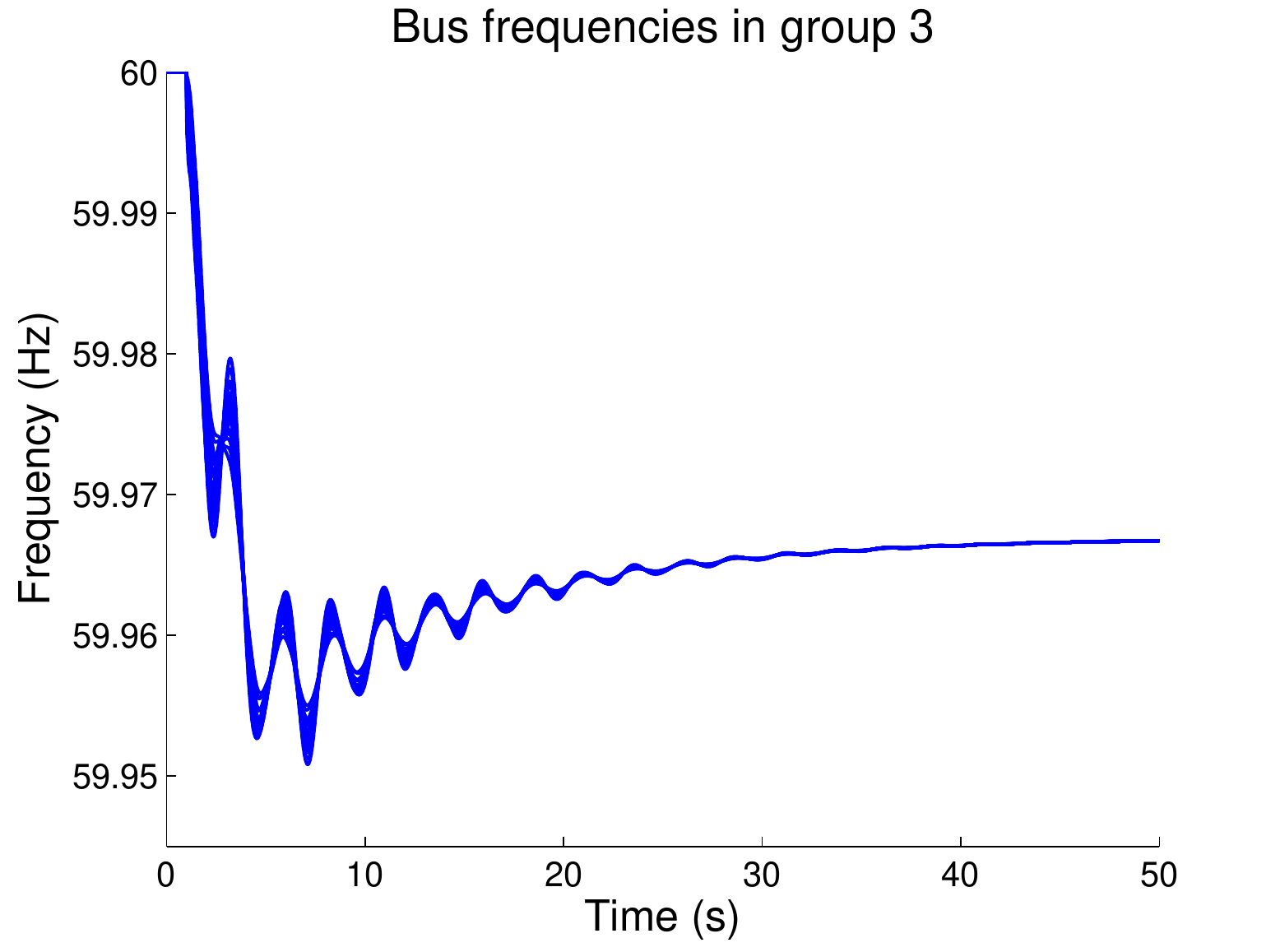}\label{fig:group_freq_3}}
\hfil
\subfigure[]
{\includegraphics[height=5.0 cm]{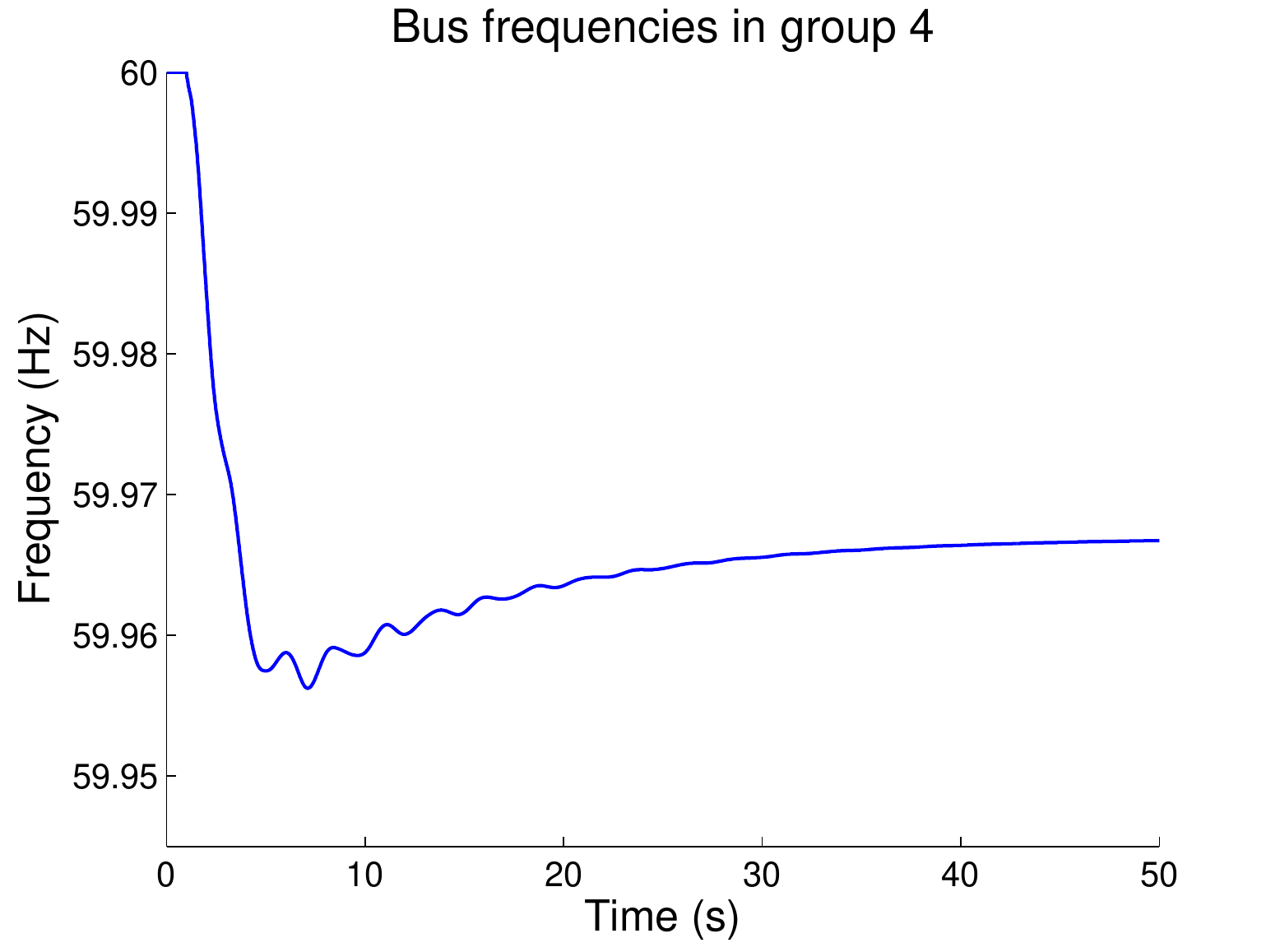}\label{fig:group_freq_4}}
\caption{Frequencies at all the 68 buses shown in four groups, without OLC.}
\label{fig:group_freq}
\end{figure*}    
We see that, during transient, the frequencies at buses within the same group are almost identical, 
but the frequencies at buses from different groups are quite different.  Moreover the time it takes for 
these different frequencies to converge to a common system frequency is on the same order as the 
time for these frequencies to reach their (common) equilibrium value.

\section{Proofs of Lemmas}\label{sec:proofs}

\subsubsection{Proof of Lemma \ref{lemma.1}}
\label{subsec:prooflemma1}

From \eqref{eq:pvar} either $c_j'(d_j(\nu))=\nu$ or $d_j'(\nu)=0$, and hence in \eqref{eq:defPhi} we have
\IEEEbq\nonumber
&&\frac{d }{d\nu}\left(c_j(d_j(\nu))-\nu d_j(\nu)\right)  
\\ &=& c_j'(d_j(\nu))d_j'(\nu)-d_j(\nu)-\nu d_j'(\nu)=-d_j(\nu) \nonumber
\IEEEeq
and therefore
\IEEEbq
\frac{\partial \Phi}{\partial \nu_j}\left(\nu\right)   =  \Phi_j'(\nu_j) =  
	-  d_j(\nu_j) - D_j \nu_j + P_j^m. \nonumber
\IEEEeq
Hence the Hessian of  $\Phi$ is diagonal.  Moreover,
since $d_j(\nu_j)$ defined in \eqref{eq:pvar} is nondecreasing in $\nu_j$, we have 
\IEEEbq
\frac{\partial^2 \Phi}{\partial \nu_j^2}\left(\nu \right)   =  \Phi_j''(\nu_j)  =  
- d_j'(\nu_j) - D_j  <  0 \nonumber
\IEEEeq
and therefore $\Phi$ is strictly concave over $\mathbb{R}^{|\node|}$.
\qed

\subsubsection{Proof of Lemma \ref{lemma.2}}
\label{subsec:prooflemma2}
Let $g$ denote the objective function of OLC with the domain $\mathcal{D} := \left[\underline d_1, \overline d_1\right] \times\dots\times\left[\underline d_{|\node|}, \overline d_{|\node|}\right]\times\mathbb{R}^{|\node|}$. 
Since $c_j$ is continuous on $\left[\underline{d}_j, \overline{d}_j\right]$, $\sum_j  c_j(d_j)$ is lower
bounded, i.e., $\sum_j  c_j(d_j) > \underline{C}$ for some $\underline{C}>-\infty$.
Let $(d', \hat{d}')$ be a feasible point of OLC (which exists by Condition \ref{cond.1}). Define the set $\mathcal{D}' := \left\{(d,\hat d) \in \mathcal{D} \left|\right. \hat{d}_j^2 \leq 2 D_j \left(g(d', \hat{d}') - \underline{C}\right),~\forall  j \in \node\right\}$. Note that for any $(d,\hat d)  \in \mathcal{D}\backslash\mathcal{D}'$, there is some $i \in \node$ such that $\hat{d}_i^2 > 2 D_i \left(g(d', \hat{d}') - \underline{C}\right)$, and thus 
\bqn
g(d, \hat{d}) > \underline{C} + \frac{\hat{d}_i^2}{2D_i}  > g(d', \hat{d}').
\eqn
Hence any optimal point of OLC must lie in $\mathcal{D}'$. By Condition \ref{cond.1} the objective function $g$ of OLC is continuous and strictly convex over the compact convex set $\mathcal{D}'$, and thus has a minimum $g^* > -\infty$ attained at a unique point $(d^*, \hat{d}^*) \in \mathcal{D}'$.

Let $(d', \hat{d}') \in \mathcal{D}$ be a feasible point of OLC, then $d_j = \left(\underline d_j + \overline d_j\right)/2$, $\hat d_j = \hat d_j' - d_j + d_j'$ specify a feasible point $(d,\hat d) \in \mathbf{relint}~\mathcal{D}$, where $\mathbf{relint}$ denotes the relative interior \cite{boyd2004convex}. Moreover the only constraint of OLC is affine. Hence there is zero duality gap between OLC and its dual, and a dual optimal $\nu^*$ is attained since $g^* > -\infty$ \cite[Sec. 5.2.3]{boyd2004convex}. By Appendix \ref{subsec:prooflemma1}), $\sum_{j \in \node}  \Phi_j''(\nu) = - \sum_{j \in \node} \left(d_j'(\nu) + D_j\right) < 0 $, i.e., the objective function of the dual of OLC is strictly concave over $\mathbb{R}$, which implies
the uniqueness of $\nu^*$. Then the optimal point $(d^*, \hat d^* )$ of OLC satisfies $d_j^* = d_j(\nu^*)$ given by \eqref{eq:pvar} and $\hat{d}_j^* = D_j \nu^*$ for $j \in \node$.
\qed

\subsubsection{Proof of Lemma \ref{lemma.3}}
\label{subsec:prooflemma3}

From the proof of Lemma \ref{lemma.1}, the Hessian
$\frac{\partial^2 \tilde L}{\partial \omega_{\generator}^2} (\omega_\generator, P) =
\frac{\partial^2 \Phi_{\generator}}{\partial \omega_{\generator}^2} (\omega_{\generator})$ is diagonal and 
negative definite for all $\omega_{\generator} \in \mathbb{R}^{|\generator|}$.
Therefore $\tilde L$ is strictly concave in $\omega_{\generator}$. 
Moreover from \eqref{eq:defLtilde} and the fact that $\frac{\partial L_\load} {\partial \omega_\load} \left(\omega_{\load}(P), P\right) =0$, we have 
\IEEEbq\label{eq:prooflemma3.1}
\frac{\partial \tilde L}{\partial P}(\omega_{\generator}, P) = -\omega_{\generator}^T C_{\generator} 
 -\omega_{\load}^T(P) C_{\load}.
\IEEEeq
Therefore we have (using \eqref{eq:dwdP})
\IEEEbq
\frac{\partial^2 \tilde L}{\partial P^2}(\omega_{\generator}, P) & = &
-C_{\load}^T \ \frac{\partial \omega_{\load}} {\partial P} (P)  \nonumber 
\\
& = &
- C_{\load}^T
 \left( \frac{\partial^2 \Phi_{\load}}{\partial \omega_{\load}^2} \left(\omega_\load(P)\right) \right)^{-1} C_{\load}. \nonumber
 \IEEEeq
From  the proof of Lemma \ref{lemma.1}, $\frac{\partial^2 \Phi_{\load}}{\partial \omega_{\load}^2}$
is diagonal and negative definite.
Hence $\frac{\partial^2 \tilde L}{\partial P^2}(\omega_{\generator}, P)$ is positive semidefinite
and $\tilde L$ is convex in $P$ ($\tilde L$ may not be strictly convex in $P$ because
$C_\load$ is not necessarily of full rank).
\qed

\subsubsection{Proof of Lemma \ref{lemma.4}}
\label{subsec:prooflemma4}

The equivalence of \eqref{eq:cond.dotU0_1} and \eqref{eq:cond.dotU0_2} follows directly
from the definition of $\omega_\load(P)$. 
To prove that \eqref{eq:cond.dotU0_1} is necessary and sufficient for $\dot{U}(\omega, P)=0$,
we first claim that the discussion preceding the lemma implies that
$(\omega, P) = \left(\omega_{\generator}, \omega_{\load}, P\right)$ satisfies 
$\dot{U}(\omega, P) = 0$ if and only if 
\IEEEbq\label{eq:cond.dotU0}
\omega_{\generator} = \omega^* 1_{\generator} &\quad \text{and}\quad& \frac{\partial \tilde L} {\partial P} (\omega_{\generator}, P) \left(P -{P}^*\right) =0.
\IEEEeq
Indeed if \eqref{eq:cond.dotU0} holds then the expression in \eqref{eq:dotU.1} evaluates to zero.   Conversely, if $\dot{U}(\omega, P)=0$,
then the inequality in (\ref{eq:dotU.2}) must hold with equality, which is possible only if 
$\omega_{\generator} = \omega^* 1_{\generator} $ since  $\tilde L$ is {\em strictly} concave  in $\omega_{\generator}$. Then we must have $\frac{\partial \tilde L} {\partial P} (\omega_{\generator}, P) \left(P -{P}^*\right) =0$ since the expression in \eqref{eq:dotU.1} needs to be zero. 
Hence we only need to establish the equivalence of \eqref{eq:cond.dotU0} and \eqref{eq:cond.dotU0_1}. Indeed, with $\omega_{\generator} = \omega^* 1_{\generator}$, the other part of \eqref{eq:cond.dotU0} 
becomes
\IEEEbq
&&\frac{\partial \tilde L} {\partial P} (\omega^* 1_{\generator}, P) \left(P -{P}^*\right) 
%  & =&\left(\frac{\partial L_\generator} {\partial P} (\omega^* 1_{\generator}, P) + \frac{\partial L_\load} {\partial \omega_\load} \left(\omega_{\load}(P), P\right) \frac{\partial \omega_\load} {\partial P} (P) + \frac{\partial L_\load} {\partial P} \left(\omega_{\load}(P), P\right)  \right)\left(P -{P}^*\right) \\
   \nonumber 
\\
&=& -\left[\omega^* 1^T_{\generator} ~~\omega^T_{\load}(P)\right] C (P-P^*)\label{eq:equicond.dotU0.0}
\\
& =& -\left[0 ~~\omega^T_{\load}(P)- \omega^* 1_{\load}^T\right] C (P-P^*)\label{eq:equicond.dotU0.1} 
\\
%& =& -\left(\omega_{\load}(P)- \omega^* 1_{\load}\right)^T C_\load (P-P^*)\nonumber \\
 & =& -\left(\omega_{\load}(P)- \omega^* 1_{\load}\right)^T \left[\frac{\partial \Phi_\load}{\partial \omega_{\load}}\left(\omega_{\load}(P)\right)-\frac{\partial \Phi_\load}{\partial \omega_{\load}}(\omega^* 1_{\load})\right]^T  \nonumber 
\\* 
\label{eq:equicond.dotU0.3}
 \IEEEeq
where \eqref{eq:equicond.dotU0.0} results from \eqref{eq:prooflemma3.1}, the equality in \eqref{eq:equicond.dotU0.1} holds since $1_{\node}^T\, C=0$, and \eqref{eq:equicond.dotU0.3} results from \eqref{eq:2v} and \eqref{eq:kkt_1}. Note that $\Phi_{\load}$ is separable over $\omega_j$ for $j \in \load$ and, from \eqref{eq:defPhi}, $ \Phi_j'(\omega_j) =-  d_j(\omega_j) - D_j \omega_j + P_j^m$.
Writing $D_{\load} := \text{diag}(D_j, j \in \load)$ we have 
 \IEEEbq
 &&\frac{\partial \tilde L} {\partial P} (\omega^* 1_{\generator}, P) \left(P -{P}^*\right)  \nonumber
\\
 & =& \left(\omega_{\load}(P)- \omega^* 1_{\load}\right)^T D_{\load} \left(\omega_{\load}(P)- \omega^* 1_{\load}\right)  \nonumber \\
&&+\sum\limits_{j \in \load} \left(\omega_j(P) - \omega^*\right)\left(d_j\left(\omega_j(P)\right)-d_j(\omega^*)\right).
\label{eq:equicond.dotU0.4}
\IEEEeq
Since $d_j(\omega_j)$ defined in \eqref{eq:pvar} is nondecreasing in $\omega_j$,
each term in the summation above is nonnegative for all $P$.
Hence \eqref{eq:equicond.dotU0.4} evaluates to zero if and only if 
$\omega_\load(P)= \omega^* 1_{\load}$, establishing the equivalence between \eqref{eq:cond.dotU0} and \eqref{eq:cond.dotU0_1}.
\qed

\subsubsection{Proof of Lemma \ref{lemma.5}}
\label{subsec:prooflemma5}
The proof of LaSalle's invariance principle in \cite[Thm. 3.4]{khalil2002nonlinear} shows that $(\omega(t), P(t))$
approaches its positive limit set $Z^+$ which is nonempty, compact, invariant and a subset of $E$, as $t \rightarrow \infty$.
It is then sufficient to show that $Z^+ \subseteq Z^*$, i.e., considering any point $(\omega, P) = \left(\omega_\generator, \omega_\load, P\right) \in Z^+$, to show that $(\omega, P) \in Z^*$. By \eqref{eq:defZstar}, \eqref{eq:defE} and the fact that $(\omega, P) \in E$, we only need to show that 
\IEEEbq
C_{\generator} P  =   \left[ \frac{\partial \Phi_{\generator}}{\partial \omega_\generator}
				\left(\omega_\generator \right) \right]^T  . \label{eq:prooflemma5.1}
\IEEEeq
%Since $(\omega, P) \in E$, we have $\omega_\generator= \omega^* 1_{\generator}$
%by Lemma \ref{lemma.4}
Since $Z^+$ is invariant with respect to  \eqref{eq:1v}--\eqref{eq:3v}, 
a trajectory $(\omega(t), P(t))$ that starts in $Z^+$ must stay in $Z^+$, and hence stay in $E$. By \eqref{eq:defE}, $\omega_\generator(t) = \omega^* 1_{\generator}$ for all $t \geq 0$, and therefore 
$\dot{\omega_\generator}(t) = 0$ for all $t \geq 0$.  Hence by \eqref{eq:1v} any trajectory $\left(\omega(t), P(t)\right)$ in $Z^+$ must satisfy
\IEEEbq
C_{\generator} P(t) & = &  \left[ \frac{\partial \Phi_{\generator}}{\partial \omega_\generator}\left(\omega_\generator(t)\right) \right]^T, \quad \forall t \geq 0 \nonumber
\IEEEeq
which implies \eqref{eq:prooflemma5.1}.
\qed

\section*{Acknowledgment}

The authors would like to thank the anonymous referees for their careful reviews and valuable comments and suggestions. They also thank Janusz Bialek, Ross Baldick, Jeremy Lin, Lang Tong, and Felix Wu for very helpful discussions on the dynamic network model, and thank Lijun Chen for discussions on the analytic approach and Alec Brooks of AeroVironment for suggestions on practical issues.

\ifCLASSOPTIONcaptionsoff
  \newpage
\fi



%\bibliographystyle{IEEEtran}


\begin{IEEEbiography}[{\includegraphics[width=1in,height=1.25in,clip,keepaspectratio]{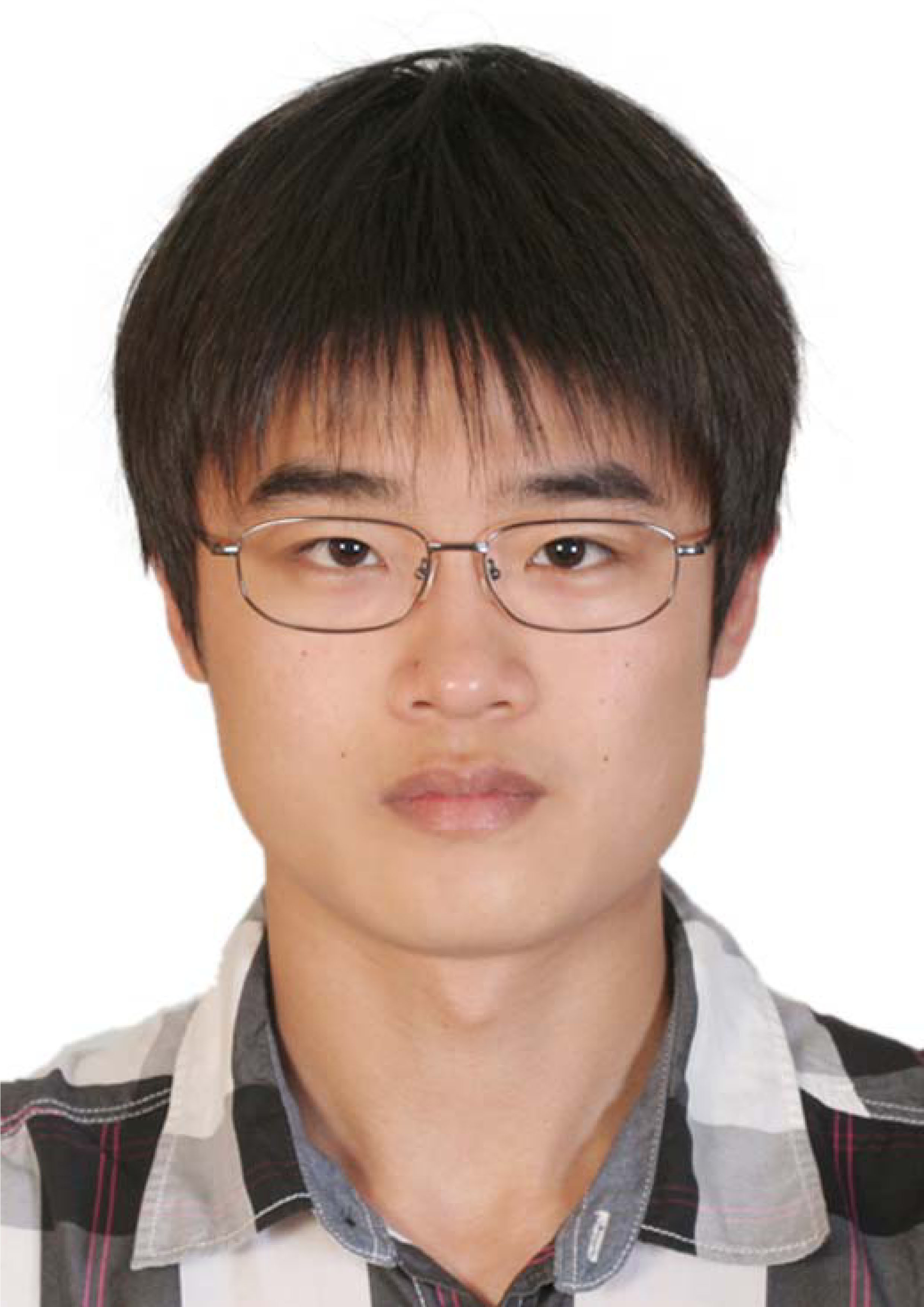}}]{Changhong Zhao} (S'12) received the B.S. degree in automatic control from Tsinghua University, Beijing, China, in 2010. 

He is currently pursuing the Ph.D. degree in electrical engineering at California Institute of Technology, Pasadena, CA, USA. His research is on the dynamics, stability, and load-side control and optimization in power systems.
\end{IEEEbiography}

\begin{IEEEbiography}[{\includegraphics[width=1in,height=1.25in,clip,keepaspectratio]{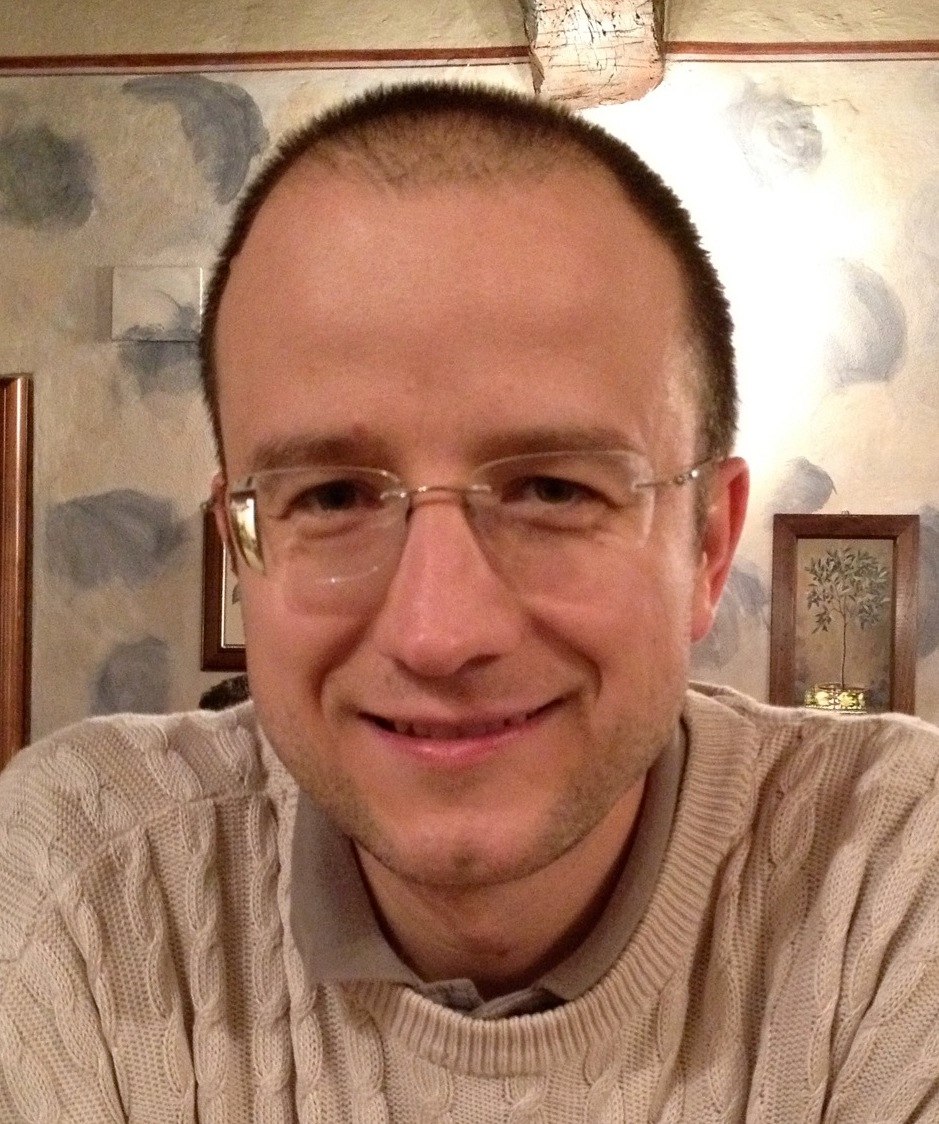}}]{Ufuk Topcu} (M'08) received the Ph.D. degree from the University of California, Berkeley, USA, in 2008. 

He is a Research Assistant Professor at the University of Pennsylvania, Philadelphia, PA, USA. His research is on the analysis, design, and veriﬁcation of networked, information-based systems with projects in autonomy, advanced air vehicle architectures, and energy networks. He was a Postdoctoral Scholar at California Institute of Technology, Pasadena, CA, USA, between 2008 and 2012.
\end{IEEEbiography}

\begin{IEEEbiography}[{\includegraphics[width=1in,height=1.25in,clip,keepaspectratio]{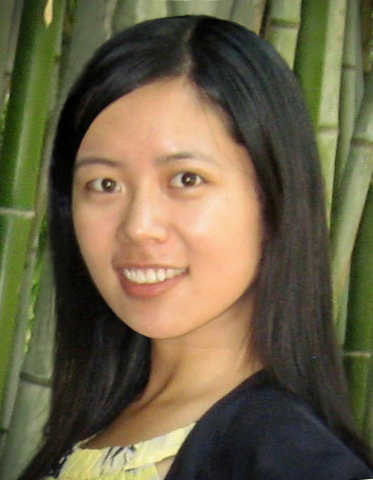}}]{Na Li} (M'13) received the B.S. degree in mathematics from ZheJiang University, China, in 2007, and the Ph.D. degree in control and dynamical systems from California Institute of Technology, Pasadena, CA, USA, in 2013.

She is currently a Postdoctoral Associate of the Laboratory for Information and Decision Systems at Massachusetts Institute of Technology, Cambridge, MA. Her research is on power and energy networks, systems biology and physiology, optimization, game theory, decentralized control and dynamical systems. She entered the Best Student Paper Award finalist in the \emph{2011 IEEE Conference on Decision and Control}.
\end{IEEEbiography} 

\begin{IEEEbiography}[{\includegraphics[width=1in,height=1.25in,clip,keepaspectratio]{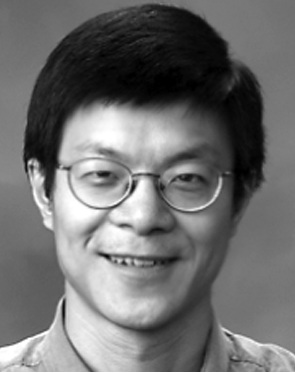}}]{Steven Low} (F'08) received the B.S. degree from Cornell University, Ithaca, NY, USA, and the Ph.D. degree from the University of California, Berkeley, USA, both in electrical engineering.

He is a Professor of the Computing and Mathematical Sciences and Electrical Engineering Departments at the California Institute of Technology, Pasadena, CA, USA. Before that, he was with AT$\text{\&}$T Bell Laboratories, Murray Hill, NJ, USA, and the University of Melbourne, Australia.
He is a Senior Editor of the \textsc{IEEE Journal on Selected Areas in Communications} (and the mentor for the annual JSAC series on Smart Grid), a Senior Editor of the \textsc{IEEE Transactions on Control of Network Systems},
a Steering Committee Member of the \textsc{IEEE Transactions on Network Science and Engineering},
and on the editorial board of \emph{NOW Foundations and Trends in Networking}, and in \emph{Power Systems}.
He also served on the editorial boards of \textsc{IEEE/ACM Transactions on Networking}, \textsc{IEEE Transactions on Automatic Control}, \emph{ACM Computing Surveys}, \emph{Computer Networks Journal}.
\end{IEEEbiography}


%\vfill


\end{document}